\documentclass[pdflatex,prd,twocolumn,showpacs,showkeys,superscriptaddress,nofootinbib]{revtex4-2}

\usepackage{bm,amsmath,amssymb}
\usepackage{graphicx}
\usepackage{color}
\usepackage{slashed}
\usepackage{tabularx}
\usepackage{bbold}
\usepackage{bbm}
\usepackage{placeins}
\usepackage[normalem]{ulem}

\usepackage[colorlinks=true]{hyperref}

\DeclareMathOperator\artanh{artanh}

\DeclareMathOperator\sign{sgn}

\DeclareMathOperator*{\SumInt}{%
\mathchoice%
  {\ooalign{$\displaystyle\sum$\cr\hidewidth$\displaystyle\int$\hidewidth\cr}}
  {\ooalign{\raisebox{.14\height}{\scalebox{.7}{$\textstyle\sum$}}\cr\hidewidth$\textstyle\int$\hidewidth\cr}}
  {\ooalign{\raisebox{.2\height}{\scalebox{.6}{$\scriptstyle\sum$}}\cr$\scriptstyle\int$\cr}}
  {\ooalign{\raisebox{.2\height}{\scalebox{.6}{$\scriptstyle\sum$}}\cr$\scriptstyle\int$\cr}}
}

\newcommand{\Tr}{\mathrm{Tr}}

\newcommand{\Imag}{{\text{Im}\,}}
\newcommand{\Real}{{\text{Re}\,}}

\newcolumntype{L}[1]{>{\raggedright\arraybackslash}p{#1}} 
\newcolumntype{C}[1]{>{\centering\arraybackslash}p{#1}} 
\newcolumntype{R}[1]{>{\raggedleft\arraybackslash}p{#1}} 

\graphicspath{{.}{figures/}}

\begin{document}

\title{\boldmath Self-consistent $O(4)$ model spectral functions from analytically continued FRG flows}

\newcommand{\KFU}{Institut f\"ur Physik, Karl-Franzens-Universit\"at Graz, NAWI Graz, Universit\"atsplatz~5, 8010 Graz, Austria}
\newcommand{\JLU}{Institut f\"ur Theoretische Physik, Justus-Liebig-Universit\"at Giessen, Heinrich-Buff-Ring 16, 35392 Giessen, Germany}
\newcommand{\JLUEXP}{II. Physikalisches Institut, Justus-Liebig-Universit\"at Giessen, Heinrich-Buff-Ring 16, 35392 Giessen, Germany}
\newcommand{\HFHF}{Helmholtz Forschungsakademie Hessen f\"ur FAIR (HFHF), Campus Giessen, 35392 Giessen, Germany}

\author{Christopher Jung}\affiliation{\JLU}
\author{Jan-Hendrik Otto}\affiliation{\JLUEXP}
\author{Ralf-Arno Tripolt}\affiliation{\JLU} \affiliation{\KFU}
\author{Lorenz von Smekal}\affiliation{\JLU}\affiliation{\HFHF}

\begin{abstract}
In this paper we  explore practicable ways for self-consistent calculations of spectral functions from analytically continued functional renormalization group (aFRG) flow equations. As a particularly straightforward one we propose to include parametrizations of self-energies based on explicit analytic one-loop  expressions. To exemplify this scheme we calculate the spectral functions of pion and sigma meson of the $O(4)$ model at vanishing temperature in the broken phase. Comparing the results with those from previous aFRG calculations, we explicitly demonstrate how self-consistency at all momenta fixes the tight relation between particle masses and decay thresholds. In addition, the two-point functions from our new semi-analytic FRG scheme have the desired domain of holomorphy built in and can readily be studied in the entire cut-complex frequency plane, on physical as well as other Riemann sheets. This is very illustrative and allows, for example, to trace the flow of the resonance pole of the sigma meson across an unphysical sheet. In order to assess the limitations due to the underlying one-loop structure, we also introduce a fully self-consistent numerical scheme based on spectral representations with scale-dependent spectral functions. The most notable improvement of this numerically involved calculation is that it describes the three-particle resonance decay of an off-shell pion, $\pi^* \to \sigma\pi\to3\pi$. Apart from this further conceptual improvement, overall agreement with the results from the considerably simpler semi-analytic one-loop scheme is very encouraging, however. The latter can therefore provide a sound and practicable basis for  self-consistent calculations of spectral functions in more realistic effective theories for warm and dense matter.
\end{abstract}



\maketitle

\section{Introduction}
The properties of strong-interaction matter, i.e.~matter subject to the strong force described by Quantum Chromodynamics (QCD), are in the focus of intense experimental and theoretical study worldwide. In particular at finite temperature and high baryon density, where one searches for a QCD critical endpoint \cite{Akiba:2015jwa,Bluhm:2020mpc}, this is a very challenging task. 
In the region of the QCD phase diagram relevant for the equation of state in the interior of neutron stars or their merger events neither lattice QCD, perturbation theory, nor chiral perturbation theory can be applied for different reasons without restrictions. In this regime, functional approaches like the Functional Renormalization Group (FRG) or Dyson-Schwinger Equations (DSEs) might help because they can in principle be applied at any temperatures and baryon chemical potential, see e.g.~\cite{Fischer:2018sdj,Gao:2020fbl} for recent compilations of results on the QCD phase diagram. Especially the FRG approach is particularly well suited to describe equilibrium thermodynamics and phase transitions. In order to avoid the peculiarities associated with gauge invariance and confinement in QCD, one thereby often employs universality and structurally simpler replacement theories to study static and dynamic critical phenomena. 

However, these functional approaches are originally also formulated in Euclidean spacetime which hampers the access to real-time quantities like spectral functions or transport coefficients due to the analytic continuation problem. In relativistic theories this entails a transition from Euclidean to Minkowski spacetime, for various reconstruction methods to achieve that see for example \cite{Vidberg:1977,Jarrell:1996rrw,Asakawa:2000tr,Mueller:2010ah,Burnier:2013nla,Fischer:2017kbq,Tripolt:2018xeo}. This is another challenging problem and, when based on a finite number of discrete data points with errors, even an ill-posed numerical problem. 

For ab-initio simulations of the static and dynamic universal critical behavior of spectral functions near continuous thermal phase transitions, one can resort to classical-statistical field theory \cite{Aarts:2001yx,Berges:2009jz,Schlichting:2019tbr,Schweitzer:2020noq}.
Reconstructing spectral functions from Euclidean data can in principle also be avoided in  non-perturbative functional methods, e.g.~based on the 2-PI effective action \cite{Berges:2004yj,Berges:2005hc,Roder:2005vt,Shen:2020jya} or Dyson-Schwinger equations \cite{Fischer:2020xnb,Horak:2020eng}.
Within the FRG, several techniques have been developed recently that are based on a direct computation of correlation functions in Minkowski spacetime and hence avoid the need for any numerical reconstruction method \cite{Strodthoff:2011tz,Floerchinger:2011sc,Kamikado:2012bt,Mesterhazy:2015uja,Huelsmann:2020xcy,Tan:2021zid}. 

The first FRG calculation of the $O(4)$ model spectral functions for pions and the sigma meson that was based on a direct analytic continuation of the Euclidean FRG flow equations for the corresponding two-point functions was presented in \cite{Kamikado:2013sia}. The calculation was set up in a thermodynamically consistent way, in that the flow of the two-point functions for vanishing external momenta, at fixed field values agreed with the corresponding flow for the curvature of the effective potential at the same values. However, the pion and sigma propagators inside these analytically continued FRG (aFRG) flows were still the simple quasi-particle constituent propagators of the leading order derivative expansion, i.e.~the local potential approximation (LPA), that was used to calculate the effective potential. This means that the two-point functions inside the flows only contained the scale-dependent quasi-particle masses but no momentum-dependent self-energy corrections. As a result of this missing self-consistency beyond the zero-momentum limit, a certain mismatch arose between the physical pole masses and the location of decay thresholds in the calculated spectral functions, for example. 

Some of these shortcomings were overcome in \cite{Strodthoff:2016pxx} where a self-consistent approach for the calculation of real-time correlation functions was proposed and applied to the $O(4)$ model in the vacuum. Therein, the analytic continuation was performed as a purely numerical procedure put forward in \cite{Pawlowski:2015mia}. This procedure was, however, found to be very computation time intensive and results for the spectral functions were only obtained up to rather low energies. In addition, access to the analytic structure of the real-time correlation functions was limited. The numerical analytic continuation procedure of \cite{Pawlowski:2015mia} was extended to finite temperature and finite spatial momentum in \cite{Pawlowski:2017gxj} where the spectral functions were obtained for a larger range of energies, although not in a fully self-consistent way.

In this work we further develop and study self-consistent approaches to calculate real-time correlation and spectral functions. A particularly 
straightforward one is based on the idea \cite{Floerchinger:2011sc} of  including parametrized self-energies with the required analyticity properties directly in the ansatz for the effective average action and calculate the flows of all parameters. The difficulty then is to find such parametrizations with the correct domain of holomorphy for two-point functions in local quantum field theory which is not entirely straightforward in general. Here, we achieve this by simply using explicit analytic expressions of one-loop form to parametrize the self-energies of pion and sigma meson at zero temperature with broken $O(4)$ symmetry. Because the resulting parametrizations are known analytically, on all Riemann sheets in fact, the analytic continuation from Euclidean to Minkowski spacetime is then straightforward again, 
and gives immediate access to the spectral functions as well as to the real-time propagators at arbitrary energies. At the same time the analytic expressions for the parametrized self-energies are also used in the propagators inside the flow of the effective potential which thus includes the same non-trivial momentum dependence  of the two-point functions as encoded in the spectral functions in a self-consistent manner. 

Comparing our results for the $O(4)$ model spectral functions from this new self-consistent scheme with those from a corresponding aFRG calculation that is built on the LPA truncation as in most previous aFRG applications, see e.g.~\cite{Tripolt:2013jra, Tripolt:2014wra, Yokota:2016kyz, Tripolt:2016cey, Yokota:2017uzu, Jung:2016yxl,Wang:2017vis, Wang:2018osm, Tripolt:2020irx}, we explicitly demonstrate that the mismatch between pole masses and decay thresholds is thereby removed. In addition, the effects of the momentum-dependent self energies are now self-consistently included in the calculation of the effective potential. Moreover, the analytic structure of the propagators is known which gives access to the spectral functions as well as to the propagators in the cut-complex frequency plane, on the physical and other Riemann sheets. 

After having established the merits of these self-consistent flows with parametrized self-energies, called the `SC1L' flows below, we also assess their limitations. These clearly are inherited from the explicit one-loop forms of the self-energies used here, which limit the structure of the possible imaginary parts from Cutkosky's cutting rules. We therefore also perform fully self-consistent aFRG calculations for comparison, with numerically computed self-energies included. This is possible based on spectral representations with scale-dependent spectral functions that are recomputed with every integration step in the FRG flow as we will explain. As a proof of principle, we demonstrate that we are able for the first time to describe the three particle resonance decay contribution to the pion spectral function, of an off-shell pion into three pions via a broad sigma resonance, at the correct threshold of three times the pion mass. This clearly comes from an imaginary part  beyond one-loop. At the same time, the overall modifications in the fully self-consistent numerical calculation, as compared to the semi-analytic SC1L flows, are otherwise surprisingly small.


This paper is organized as follows: In Sec.~\ref{sec:theoretical_setup} we start 
with a brief introduction of the general FRG concepts needed here. In Sec.~\ref{sec:LPA_setup} we summarize the analytically continued LPA setup which will be used for comparison with the self-consistent SC1L flows introduced in Sec.~\ref{sec:one_loop_setup}. In Sec.~\ref{sec:UV_and_numerics} we discuss parameter sets, UV conditions, and our numerical implementation. In Sec.~\ref{sec:pot_couplings_masses} we compare the results for effective potential,  particle masses, and the coupling constants from LPA and SC1L flows. The corresponding  results for the two-point functions in Euclidean and Minkowski spacetime, from analytic continuation, together with the spectral functions are compared  in Sec.~\ref{sec:2PF_and_SF}. In Sec.~\ref{sec:analytic_structure} we discuss the analytic structure of the SC1L two-point functions in the complex-momentum plane. As a first step to include self-energies beyond one-loop structure we describe a fully self-consistent numerical framework, based on spectral representations with scale dependent spectral functions, in Sec.~\ref{sec:sc_spectral_functions}, where we also compare the fully self-consistent with the SC1L results from the previous sections to assess the overall quality of the latter. We close with our summary and an outlook in Sec.~\ref{sec:summary}. Details on the explicit analytic expressions for the self-energy parametrizations and loop functions are provided together with other technicalities in several appendices.

\section{Theoretical setup}
\label{sec:theoretical_setup}
The FRG is a powerful and versatile non-perturbative framework that aims at calculating the effective action by successively integrating out quantum fluctuations in momentum space, for reviews see for example \cite{Berges:2000ew,Polonyi:2001se,Pawlowski:2005xe,Schaefer:2006sr,Kopietz:2010zz,Braun:2011pp, Friman:2011zz, Gies:2006wv,Dupuis:2020fhh}. In the formulation pioneered by C.~Wetterich \cite{Wetterich:1992yh}, the RG-scale dependence of the effective average action $\Gamma_k$, which interpolates between the bare action $S\simeq \Gamma_\Lambda$ at an ultraviolet (UV) scale $\Lambda$ and the full quantum effective action $\Gamma\equiv\Gamma_{k=0}$ in the infrared (IR), is given by the following flow equation,
\begin{align}
\label{eq:Wetterich_equation}
\partial_k\Gamma_k[\phi]=\frac{1}{2}\Tr{\frac{\partial_k R_k}{\Gamma_k^{(2)}[\phi]+R_k}}.
\end{align}
Therein, the regulator function $R_k$ acts as a mass term and suppresses fluctuations of low-momentum modes, $p \lesssim k$, while the high-momentum modes, $p \gtrsim k$, are already integrated out and included in $\Gamma_k$. For a discussion of how to devise optimized regulators in a particular truncation where this can be quite non-trivial, see \cite{Pawlowski:2015mlf}. Apart from the regulator function, the Wetterich equation only depends on the second functional field derivative of the effective action, which is denoted as $\Gamma_k^{(2)}[\phi]$. The trace in Eq.~(\ref{eq:Wetterich_equation}) represents a summation over internal indices as well as an integration over momentum space, which gives rise to a simple one-loop structure of the Wetterich equation since $(\Gamma_k^{(2)}[\phi]+R_k)^{-1}$ represents the full scale-dependent propagator $D_k$. 

In general, the two-point function $\Gamma_k^{(2)}[\phi]$ has to be obtained from its respective flow equation, which is given by the second functional derivative of Eq.~(\ref{eq:Wetterich_equation}). This flow equation in turn depends on the three- and four-point functions, $\Gamma_k^{(3)}[\phi]$ and $\Gamma_k^{(4)}[\phi]$, for which additional flow equations have to be derived and solved. This leads to an infinite tower of equations which has to be truncated at some point. To achieve this, one may chose a suitable ansatz for the functional form of the effective action. Standard expansion schemes for the effective action are for example given by vertex expansions or a derivative expansion, see e.g.~\cite{Berges:2000ew}. While vertex expansions allow to take into account the full momentum dependence of $n$-point correlation functions, the derivative expansion can resolve the full field dependence of the corresponding terms in the effective action already at the lowest order, i.e.~in the local potential approximation, see e.g.~\cite{Litim:2001dt,Braun:2009si}.

In this paper we use an ansatz for the effective average action which combines both aspects. It includes two-point functions containing arbitrarily many derivatives which are one-loop exact and self-consistent. All three and higher $n$-point functions are included as the scale-dependent but structure-less vertices obtained from the LPA. In a first exploratory study we here apply this scheme to the relativistic $O(4)$ model in four space-time dimensions, where our ansatz for the effective average action is then of the form:
\begin{align}
\label{eq:Ansatz_Gamma}
\Gamma_k[\phi]=&\int d^4x \left(\frac{1}{2}(\partial_\mu \phi)^2+U_k(\phi^2)-c\sigma \right)\nonumber\\
& - \frac{1}{2} \int d^4x d^4y\,
\, \phi(x)  \Sigma_k(x-y;\sigma) \phi(y) \, .  
\end{align}
Here, $U_k(\phi^2)$ is the $O(4)$ symmetric effective potential that includes the mass term and all higher-order local mesonic self-interactions, i.e.~the scale dependent $n$-point vertices with $n\ge 3$.
The $-c\sigma$ term is included to account for the explicit symmetry breaking. 
The model serves as a low-energy effective description of the light meson sector in QCD with $N_f=2$ flavors. The field $\phi$ then encodes the sigma and the pion fields, $\phi=(\sigma,\vec{\pi})^T$, and the term $c\sigma$ resembles the current quark masses in QCD that explicitly break the corresponding chiral symmetry.

The self-energies $\Sigma_k(x) $ for pions and the sigma meson are diagonal in field space, of one-loop structure, and renormalized by zero-momentum subtraction so that they vanish for constant fields in order not to interfere with the effective potential and to avoid double counting. To be precise, this implies that their two lowest moments vanish,
\begin{align}
   \int d^4x \, \Sigma_k(x;\sigma) = \int d^4x \, x\, \Sigma_k(x;\sigma) = 0\, .
\end{align}
To achieve this, the corresponding self-energies in momentum space, referred to as polarization functions $\Pi_k$ below, 
\begin{align}
  \Pi_k(p) = \int d^4x \, e^{-ipx} \,  \Sigma_k(x;\sigma) \, ,
\end{align}
are renormalized by subtraction at zero momentum so that  $\Pi_k(0) = 0$ with $\Pi_k(p)\sim p^2 $ for $p^2\to 0$. 

To systematically extend the vertex-expansion scheme, at the next order one might include self-consistent self energies of two-loop structure in combination with a correspondingly soft-momentum subtracted three-point function of one-loop form, supplemented by local $n$-point vertices for $n \ge 4$, and so forth. While straightforward in principle, these higher order expansions will eventually have to rely on numerical methods to evaluate the successively higher loop contributions to the lower $n$-point functions maintained in the expansion. At the present one-loop order we have the luxury to be able to use explicit and analytic expressions for the self-energies in the entire cut-complex plane of their analytically continued invariant momentum $p^2$ argument. This makes the analytic continuation of the flow equations for the two-point functions, to calculate the spectral functions in our aFRG framework, particularly transparent, and it therefore provides an illustrative and valuable first step. 
For constant fields, our ansatz reduces to the local potential approximation by design, and the Wetterich equation reduces to a flow equation for the effective potential $U_k(\phi^2)$ as the only scale-dependent quantity on the left hand side,
\begin{align}
\label{eq:Flow_Potential}
\partial_k U_k=\frac{1}{2}\int_q \left(
\frac{\partial_k R_k(q)}{\Gamma_{\sigma,k}^{(2)}(q)+R_k(q)}
+
3\, \frac{\partial_k R_k(q)}{\Gamma_{\pi,k}^{(2)}(q)+R_k(q)}
\right),
\end{align}
where the integration is over four-dimensional momentum space. The two-point functions on the right hand side, on the other hand, will now include the sigma and pion polarization functions $\Pi_{\sigma,k}(q)$ and $\Pi_{\pi,k}(q)$ as detailed in Sec.~\ref{sec:one_loop_setup} below. For a better comparison, however, we will first briefly review the standard LPA procedure to calculate pion and sigma spectral functions in the next subsection.

\subsection{Analytically continued LPA flows}
\label{sec:LPA_setup}
The LPA setup presented in this section was first proposed in \cite{Kamikado:2013sia} and will be briefly summarized here. The standard LPA form of the two-point functions can be obtained by taking two functional derivatives of the ansatz for the effective action, Eq.~(\ref{eq:Ansatz_Gamma}), here without the subtracted self-energies, which then simply yields
\begin{align}
\label{eq:LPA_2PF}
\Gamma_{\sigma,k}^{(2),\text{LPA}}(p)&=p^2+(m_{\sigma,k}^{c})^2,\\
\Gamma_{\pi,k}^{(2),\text{LPA}}(p)&=p^2+(m_{\pi,k}^{c})^2,
\label{eq:LPA_2PF_2}
\end{align}
where the squares of the (Euclidean or curvature) masses are given by,
\begin{align}
\label{eq:curvature_masses}
(m_{\sigma,k}^{c})^2&=2\,U_k'(\rho)+4\,\rho\, U_k''(\rho),\\
\label{eq:curvature_masses_2}
(m_{\pi,k}^{c})^2&=2\,U_k'(\rho),
\end{align}
with $\rho=\phi^2$. These two-point functions are then inserted into the flow equation for the effective potential, Eq.~(\ref{eq:Flow_Potential}), which leaves to specify the regulator function $R_k$. In order to reproduce the previous LPA setup for the $O(4)$ model spectral functions from the aFRG flows in \cite{Kamikado:2013sia} as our benchmark calculation we will here also use the three-momentum analogue of the LPA-optimized regulator \cite{Litim:2001up} from that previous study, which is given by
\begin{align}
\label{eq:regulator_3d}
R_k(\vec{p})=(k^2-\vec{p}^{\,2})\Theta(k^2-\vec{p}^{\,2}),
\end{align}
where $\Theta(x)$ is the Heaviside step function.  It only regulates spatial momenta but not the energy components at the expense of some breaking of the Euclidean $O(4)$ symmetry. This breaking was assessed and found to be negligible for external momenta well below the UV cutoff scale $\Lambda$ in the Euclidean two-point functions, with still reasonably small and only quantitative effects in the time-like domain after analytic continuation \cite{Kamikado:2013sia}. While this can be avoided in principle \cite{Pawlowski:2015mia,Pawlowski:2017gxj}, the three-dimensional regulators allow to perform the integration over the internal energy component or the corresponding Matsubara sum at finite temperature analytically which tremendously simplifies the analytic continuation procedure.

\begin{figure}[t]
	\includegraphics[width=0.49\textwidth]{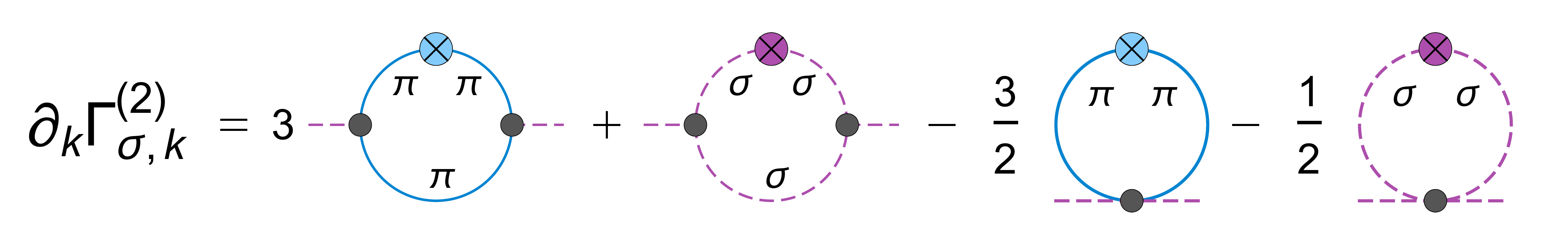}
	\includegraphics[width=0.49\textwidth]{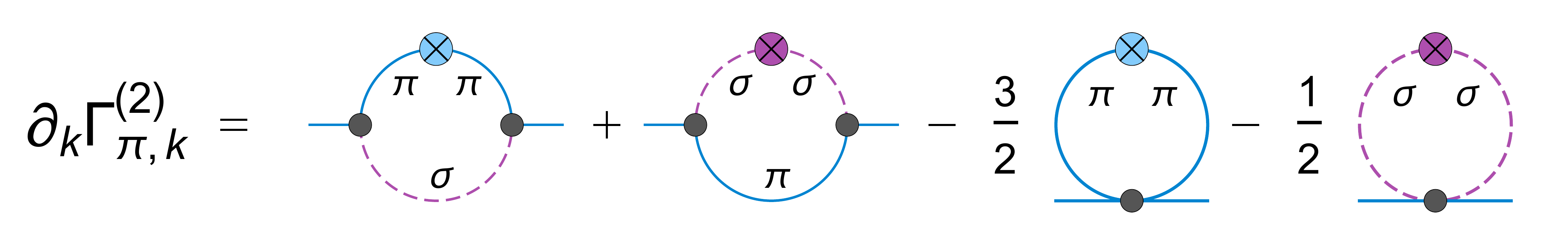}
	\caption{Diagrammatic representation of the flow equations for the sigma and the pion two-point function. Solid blue lines represent full scale-dependent pion propagators while dashed purple lines represent the sigma propagators. The crossed circle represents the derivative of the regulator function $\partial_k R_k$.}
	\label{fig:2PF_diagrams}
\end{figure}

The scale-dependent effective potential is then used as input in the flow equations for the two-point functions. These are obtained from the Wetterich equation by taking two functional field derivatives, leading to
\begin{align}
\label{eq:Gamma2_sigma}
\partial_k \Gamma_{\sigma,k}^{(2)}(p)=
&
\,3(\Gamma^{(3)}_{\sigma\pi,k})^2 J_{\pi,k}(p)
+(\Gamma^{(3)}_{\sigma,k})^2 J_{\sigma,k}(p)
\nonumber \\
&
-\frac{3}{2}(\Gamma^{(4)}_{\sigma\pi,k})^2 I^{(2)}_{\pi,k}
-\frac{1}{2}(\Gamma^{(4)}_{\sigma,k})^2 I^{(2)}_{\sigma,k},\\
\label{eq:Gamma2_pion}
\partial_k \Gamma_{\pi,k}^{(2)}(p)=
&
\,(\Gamma^{(3)}_{\sigma\pi,k})^2 J_{\pi\sigma,k}(p)
+(\Gamma^{(3)}_{\sigma\pi,k})^2 J_{\sigma\pi,k}(p)
\nonumber \\
&
-\frac{3}{2}(\Gamma^{(4)}_{\pi,k})^2 I^{(2)}_{\pi,k}
-\frac{1}{2}(\Gamma^{(4)}_{\sigma\pi,k})^2 I^{(2)}_{\sigma,k}
,
\end{align}
see Fig.~\ref{fig:2PF_diagrams} for a diagrammatic representation. Explicit expressions for the involved loop functions $I$ and $J$ as well as for the four-point vertex functions $\Gamma^{(4)}$ are given in App.~\ref{app:loop_functions}. As we will use the same three-point vertex functions in the one-loop setup as in the LPA setup, we discuss them in more detail here. They are obtained from appropriate derivatives of the ansatz for the effective action in Eq.~(\ref{eq:Ansatz_Gamma}) which leads to
\begin{align}
\label{eq:couplings-s}
\Gamma^{(3)}_{\sigma,k}&=
12\,\sigma\,U_k''(\rho)
+8\,\sigma^3\,U_k'''(\rho),\\
\Gamma^{(3)}_{\sigma\pi,k}&=
4\,\sigma\,U_k''(\rho).
\label{eq:couplings-sp}
\end{align}
We note that this setup is thermodynamically consistent, in that the flow equations for the three-point functions $\Gamma^{(3)}(p,q)$ can be expressed in terms of derivatives of the flow equation for the effective potential when evaluated at vanishing external momenta $p=q=0$,
\begin{align}
\label{eq:3PF}
\partial_k \Gamma^{(3)}_{\sigma,k}(0,0)&=
12\,\sigma\, \partial_k U_k''(\rho)
+8\,\sigma^3\, \partial_k U_k'''(\rho),\\
\partial_k \Gamma^{(3)}_{\sigma\pi,k}(0,0)&=
4\,\sigma\, \partial_k U_k''(\rho).
\end{align}
The analogous consistency relation is known to hold for the two-point functions as well \cite{Kamikado:2013sia}.

Since the vertex functions in Eqs.~(\ref{eq:Gamma2_sigma})-(\ref{eq:Gamma2_pion}) only carry field and scale dependence, the momentum dependence is generated solely by the loop functions $J$. As shown in App.~\ref{app:loop_functions}, they carry the momentum dependence of the LPA propagators which in turn depend on the scale-dependent curvature masses in Eqs.~(\ref{eq:curvature_masses})-(\ref{eq:curvature_masses_2}). The curvature masses therefore also determine the location of thresholds in the ensuing spectral functions which leads to a certain mismatch between the physical pole masses and decay thresholds. We also note that the non-trivial two-point functions obtained from these aFRG flows are not inserted back into the flow equations here. Although these two-point functions with their non-trivial momentum dependence in general provide results beyond any finite order in the derivative expansion, the self-consistency of these flows remains restricted to the lowest order LPA level, valid for the effective potential and the zero-momentum limit of the two-point functions. Fully self-consistent aFRG calculations are possible in principle, but technically and numerically very demanding as we will demonstrate in Sec.~\ref{sec:sc_spectral_functions} below. Moreover, the resulting IR two-point functions are then only known numerically which limits access to their structure in the complex plane and on different Riemann sheets. For all these reasons, we here propose a self-consistent but equally simple and illustrative one-loop setup as described in Sec.~\ref{sec:one_loop_setup} next.

We close this section by a brief discussion of the analytic continuation of the flow equations and the calculation of spectral functions. At zero temperature, the analytic continuation simply involves replacing the Euclidean energy $p_0$ by a real frequency $\omega$ in the following way,
\begin{align}
p_0\rightarrow -i(\omega+i\varepsilon),
\label{eq:continuation}
\end{align}
where the limit $\varepsilon\rightarrow 0^+$ is implicitly assumed. This prescription yields the flow equations for the retarded two-point functions at real frequency $\omega $ which can then be solved numerically. Finally, the spectral functions are defined as the imaginary parts of the retarded propagators $D^R_k= -D_k(-i\omega^+)$, with $\omega^+=\omega+i\varepsilon$, 
and can be obtained from the corresponding real-time two-point functions (with retarded self-energies) as follows,
\begin{align}
\label{eq:spectral_function}
\rho_k(\omega)&=-\frac{1}{\pi}\text{Im}\, D_k^R(\omega) \\&=
\label{eq:spectral_function_2}
\frac{1}{\pi}\frac{\text{Im}\, \Gamma^{(2),R}_k(\omega)}{\Big(\text{Re}\, \Gamma^{(2),R}_k(\omega)\Big)^2+\Big(\text{Im} \, \Gamma^{(2),R}_k(\omega)\Big)^2} \, . 
\end{align}

\subsection{Self-consistent one-loop (SC1L) flows}
\label{sec:one_loop_setup}

In this section we present our new self-consistent one-loop FRG approach. It is based on parametrizing the two-point functions by analytic expressions for the scale-dependent self-energies $\Pi_k(p)$ obtained from one-loop formulae given in Appendix \ref{app:one_loop}. By construction, these expressions are given by analytic functions in the cut-complex $p^2$-plane and allow to extract the spectral functions from the discontinuities of the corresponding propagators along this cut. With these analytic one-loop expressions, the type of processes that can occur, from Cutkosky's rule, are therefore the same as  in the analytically continued LPA flows above. The parameters involved in these one-loop parametrizations are calculated self-consistently from the aFRG flows, however, as described below. In particular, the two-particle thresholds are then determined by the physical pole masses $m^p$ of the particles involved. This fixes a slight inconsistency of the aFRG LPA flows where these thresholds are determined by the curvature masses $m^c$, cf.~Eqs.~(\ref{eq:LPA_2PF})-(\ref{eq:LPA_2PF_2}), which represent the zero-momentum limit of the two-point functions and the curvature of the effective potential in the corresponding field direction at the same time. However, these curvature masses  will differ, in general, from the physical pole masses, especially for heavier particles where extrapolation from the pole position at $-p^2 = (m^p)^2 $ to $p^2=0$ can induce sizable corrections due to the $p^2$ dependence of the self-energies.

\begin{figure}[t]
	\includegraphics[width=0.37\textwidth]{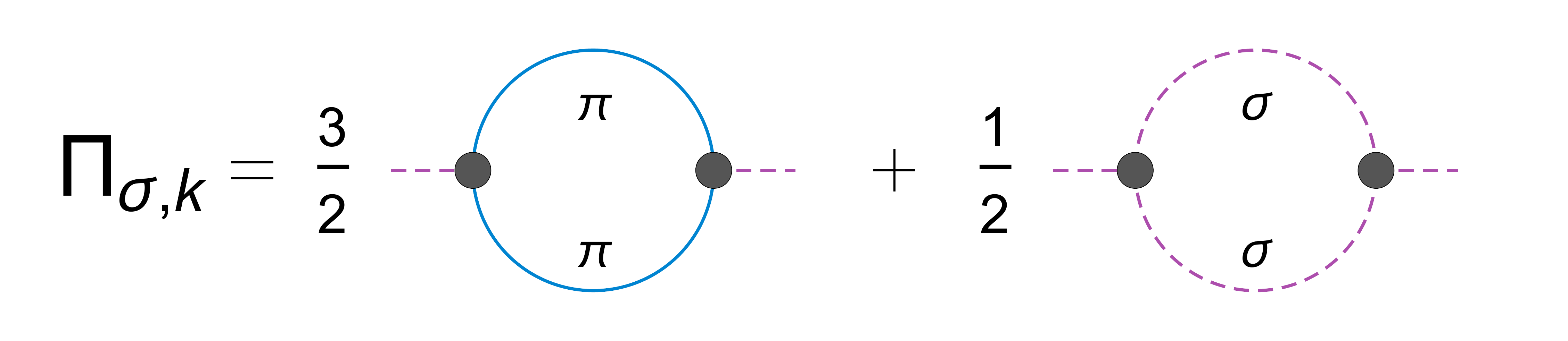}
	\includegraphics[width=0.37\textwidth]{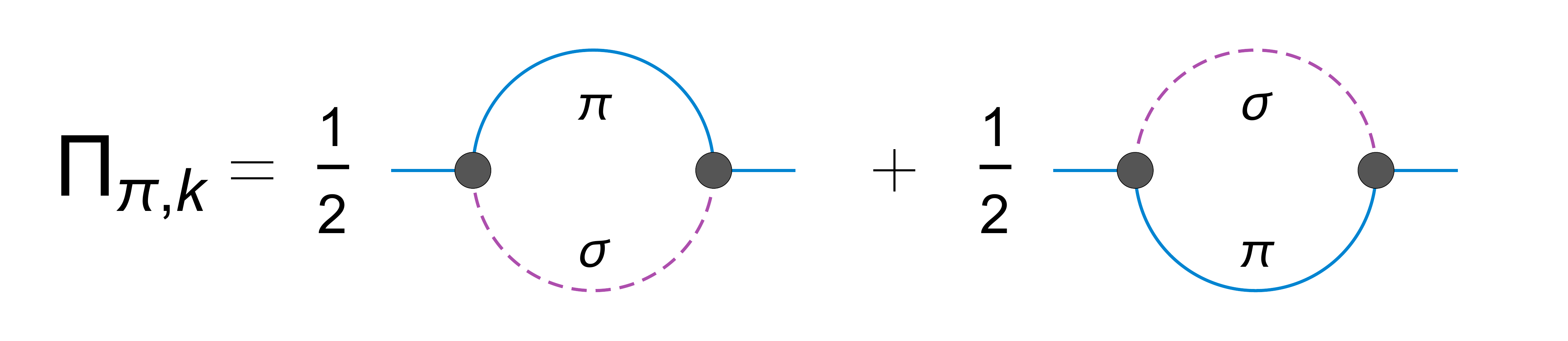}
	\caption{Diagrammatic representation of the contributions to the one-loop self energy of the sigma and the pion propagator.}
	\label{fig:1L_diagrams}
\end{figure}

With these self-energies included, the sigma and pion two-point functions are then given by
\begin{align}
\label{eq:1L_2PF}
\Gamma_{\sigma,k}^{(2),\text{1L}}(p)&=p^2+(m_{\sigma,k}^{c})^2-\Pi_{\sigma,k}(p)\, ,\\
\label{eq:1L_2PF_2}
\Gamma_{\pi,k}^{(2),\text{1L}}(p)&=p^2+(m_{\pi,k}^{c})^2-\Pi_{\pi,k}(p)\, .
\end{align}
Here, the self-energies can be split up into the contributions from the different loops, see also Fig.~\ref{fig:1L_diagrams}, 
\begin{align}
\label{eq:1L_self_energies}
\Pi_{\sigma,k}(p)&=
3\,   \frac{g^2_{\sigma\pi,k}}{16\pi^2}  \, \Pi_R(p^2, m_{\pi,k}^p, m_{\pi,k}^p) \\
&\qquad+  \frac{g^2_{\sigma,k}}{16\pi^2}  \,   \Pi_R(p^2, m_{\sigma,k}^p, m_{\sigma,k}^p)\, ,\nonumber\\
\label{eq:1L_self_energies_2}
\Pi_{\pi,k}(p)&=2\, \frac{g^2_{\sigma\pi,k}}{16\pi^2}  \, \Pi_R(p^2, m_{\sigma,k}^p, m_{\pi,k}^p)\, ,
\end{align}
where $g_{\sigma,k}$ and $g_{\sigma\pi,k}$ are the scale-dependent three-point $\sigma$ and $\sigma$-$\pi$ coupling constants, $m_{\sigma,k}^p$ and $m_{\pi,k}^p$ denote the scale-dependent pole masses. 
The renormalized self-energy integrals  $\Pi_R$ are defined in App.~\ref{app:one_loop} and include the symmetry factor of $1/2$ in the diagrams, while the factor of 3 in Eq.~(\ref{eq:1L_self_energies}) reflects the three pion fields, and the factor of 2 in Eq.~(\ref{eq:1L_self_energies_2}) comes from the degeneracy of the two lower loops in Fig.~\ref{fig:1L_diagrams}. The renormalization of the otherwise UV divergent self-energy integrals is performed by zero-momentum subtraction. Hence, these self-energies vanish at zero momentum, $\Pi_R(0) = 0$, so that the relation between zero-momentum two-point function and effective potential or curvature mass, $\Gamma_{k}^{(2),\text{1L}}(0)= (m_{k}^{c})^2$ remains unchanged by construction. 
We also note that the self-energies are analytic with $\Pi_R(p^2) = \mathcal O(p^2) $ at  $p^2=0 $ in presence of a mass gap.

For the scale-dependent coupling constants $g_{\sigma,k}$ and $g_{\sigma\pi,k}$ we will use the same expressions (\ref{eq:couplings-s}) and (\ref{eq:couplings-sp}) as in the LPA setup, i.e.
\begin{align}
\label{eq:couplings}
g_{\sigma,k} =\Gamma^{(3)}_{\sigma,k}\, , \quad \mbox{and}\;\;
g_{\sigma\pi,k} =\Gamma^{(3)}_{\sigma\pi,k} \, ,
\end{align}
but evaluated at the scale-dependent global minimum of the effective potential $\rho_{0,k}$. The effective potential is obtained from Eq.~(\ref{eq:Flow_Potential}), as in the LPA setup, but now evaluated using the one-loop two-point functions, Eqs.~(\ref{eq:1L_2PF})-(\ref{eq:1L_2PF_2}). For the regulator function we will here use the four-dimensional Litim regulator,
\begin{align}
\label{eq:regulator_4d}
R_k(p)=(k^2-p^2)\Theta(k^2-p^2).
\end{align}

Given the coupling constants $g_{\sigma,k}$ and $g_{\sigma\pi,k}$, one can immediately obtain the pole masses $m_{\sigma,k}^p$ and $m_{\pi,k}^p$. For resonances, these are of course complex valued and off the physical Riemann sheet. When the widths are sufficiently small they are close to the physical sheet, and their real parts are then very well approximated by the zero crossing of the real part of the corresponding two-point functions along the timelike real axis.
In the following, we therefore define what we mean by the pole masses from the zero crossings of the real parts of the corresponding two-point functions as our proxy for the real parts of the resonance positions, i.e.~as the scale-dependent solutions $m^p_{\sigma,k}$, $m^p_{\pi,k}$ of
\begin{align}
\label{eq:pole_masses}
(m_{\sigma,k}^p)^2 &= (m_{\sigma,k}^c)^2- \text{Re}\,\Pi_{\sigma,k}(p)\Big|_{-p^2 =(m_{\sigma,k}^p)^2}  \, ,\\
(m_{\pi,k}^p)^2 &= (m_{\pi,k}^c)^2- \text{Re}\,\Pi_{\pi,k}(p)\Big|_{-p^2 =(m_{\pi,k}^p)^2} \, ,
\label{eq:pole_masses2}
\end{align}
with Eqs.~(\ref{eq:1L_self_energies}) and (\ref{eq:1L_self_energies_2}) for the two self-energies. Note that their real parts are continuous along the cut and no $\varepsilon$-prescription is thus required here. Since both two-point functions depend on both pole masses, this coupled set of equations has to be solved simultaneously at each scale $k$. Eqs.~(\ref{eq:pole_masses})-(\ref{eq:pole_masses2}) are the self-consistency equations for the scale-dependence of the mass parameters which determine that of the resonance positions and decay thresholds at the same time in a consistent manner at the expense of an only moderate increase in numerical costs.

As the analytic form of the two-point functions is fixed by Eqs.~(\ref{eq:1L_2PF})-(\ref{eq:1L_2PF_2}), there is no need to solve flow equations for them. In fact, it can be shown that solving the flow equations for the two-point functions, Eqs.~(\ref{eq:Gamma2_sigma})-(\ref{eq:Gamma2_pion}), for classical input, i.e.~for scale-independent coupling constants and masses, exactly reproduces the one-loop result. The main effects of the LPA aFRG calculation of the two-point functions discussed in Sec.~\ref{sec:LPA_setup} are therefore expected to be included in the one-loop setup proposed here, in particular when using scale-dependent parameters obtained from the FRG in the self-consistent one-loop two-point functions. 

After solving the self-consistent system of equations for the effective potential and the two-point functions, the spectral functions can be obtained from the real-time two-point functions using Eq.~(\ref{eq:spectral_function_2}). Given the analytic structure of the two-point functions, the spectral functions can be evaluated at arbitrarily high energies. This is not possible in the LPA setup since there the UV cutoff for the RG scale also limits the energy range where the spectral functions give meaningful results, see also the results shown in Sec.~\ref{sec:2PF_and_SF}. Knowing the spectral functions at arbitrary energies also allows us to check their normalization which is given by the energy-weighted sum rule
\begin{align}
\label{eq:normalization}
\int_{0}^{\infty} \rho(\omega^2)d\omega^2=1,
\end{align}
see for example \cite{Tripolt:2018qvi} for a more detailed discussion of various sum rules for the spectral functions from aFRG flows. We have verified numerically that this sum rule is indeed satisfied, independent of the particular values used for coupling constants and masses.

Finally, we note that it is also possible to explore the analytic structure of the two-point functions in the complex-momentum plane by simply evaluating them at complex arguments. In particular, it is possible to investigate different Riemann sheets and thus to identify the complex pole of the sigma meson. For more details on how to access different Riemann sheets we refer to App.~\ref{app:one_loop} while corresponding results are presented in Sec.~\ref{sec:analytic_structure}.

\begin{table}[t]
	\centering
	\begin{tabular}{C{0.8cm}|C{1.2cm}|C{1.0cm}|C{1.3cm}||C{1.0cm}|C{0.9cm}|C{0.9cm}}
		& $b_1$ & $b_2$ & $c$  & $f_\pi$ & $m_\pi^p$ & $m_\sigma^p$  \\
		&[$\Lambda ^2$] &  & [$\Lambda^{3}$]& [MeV] & [MeV] & [MeV]
		\\
		\hline\hline
		LPA &  -0.32456  & 3.923 & 0.014 &93.0 & 135 & 352 \\
		SC1L & -0.22524  & 3.372 & 0.014 &93.0 & 135 & 324
	\end{tabular}
	\caption{Numerical values for the parameters of the effective potential in the UV for the LPA setup and the self-consistent one-loop setup (SC1L), along with the resulting values for the pole masses in the IR.}
	\label{tab:pot_params} 
\end{table}

\subsection{UV conditions and numerical implementation}
\label{sec:UV_and_numerics}
For both the LPA aFRG and the self-consistent one-loop (SC1L) setup we choose the effective potential in the UV to be of the following form,
\begin{align}
\label{eq:UV_pot}
U_\Lambda(\rho)=b_1 \rho + b_2 \rho^2,
\end{align}
while the explicit symmetry breaking term, $-c\sigma$, is added in the IR since it is a renormalization group invariant term. The parameters $b_1$, $b_2$ and $c$ are chosen such as to obtain phenomenological values for the pion decay constant as well as for the pole masses of the pion and the sigma meson, where, however, in FRG studies of the $O(4)$ model the possible range for the sigma mass is typically below its experimental value, see also \cite{Kamikado:2013sia}. Explicit numerical values for the parameters chosen in the UV and the resulting IR values are given in Tab.~\ref{tab:pot_params}, where we use $\Lambda=500$~MeV for the UV scale and $k_{\text{IR}}=40$~MeV as the IR scale. In order to obtain results for $k<k_{\text{IR}}$ we use an extrapolation. We note that the pion decay constant is identified with the global minimum of the effective potential in the IR, $\sigma_0$, which in the SC1L setup needs to be renormalized due to the non-trivial wave function renormalization factors, see Sec.~\ref{sec:pot_couplings_masses} below for a more detailed discussion.

 \begin{figure}[t]
	\includegraphics[width=0.48\textwidth]{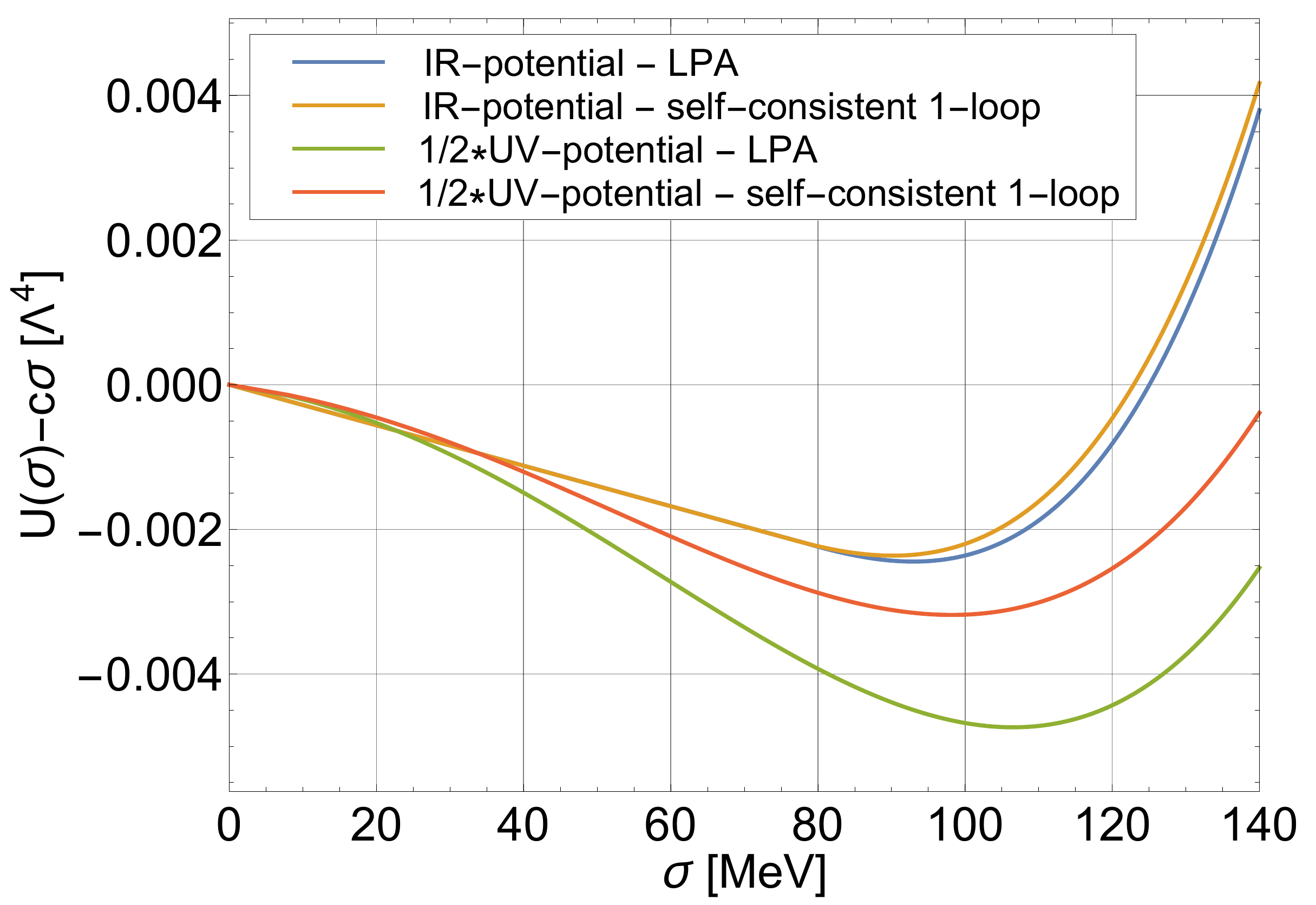}
	\caption{The effective potentials $U_k(\rho)$ obtained from the LPA flow as well as the SC1L flow are shown in the UV and in the IR, both  normalized to zero at $\sigma=0$, and with the UV potentials divided by a factor of 2 for better comparison.}
	\label{fig:pot_comparison}
\end{figure}

The flow equation for the effective potential, Eq.~(\ref{eq:Flow_Potential}), is solved numerically using the so-called grid method which is based on a discretization of the field variable $\rho$ into equidistant grid points, see for example \cite{Schaefer:2004en}. The flow equation then reduces to a set of coupled ordinary differential equations which can be solved using standard techniques where we obtain the required first and second derivatives of the effective potential using a finite-difference discretization. In the present work we used different numbers of grid points in the range $N\in [100,300]$ distributed over a $\rho$-field range of $\rho\in [0,140^2]$MeV$^2$. We note that such a finite-difference implementation has certain limitations, see e.g.~\cite{Grossi:2019urj,Grossi:2021ksl}. We therefore checked explicitly that another, more sophisticated numerical technique \cite{Koenigstein:2021syz,Steil:2021cbu}, i.e.~the Kurganov-Tadmor finite volume technique \cite{KTpaper}, yields the same numerical results.

Within the LPA setup, the solution for the scale-dependent effective potential is then used as input in the flow equations for the two-point functions, Eqs.~(\ref{eq:Gamma2_sigma})-(\ref{eq:Gamma2_pion}). In the UV, the LPA real-time two-point functions have the following form, 
\begin{align}
\label{eq:UV_Gamma2}
\Gamma^{(2)}_{\sigma,\Lambda}(\omega)&=\omega^2+\vec{p}^{\,2}+(m_{\sigma,\Lambda}^c)^2,\\
\Gamma^{(2)}_{\pi,\Lambda}(\omega)&=\omega^2+\vec{p}^{\,2}+(m_{\pi,\Lambda}^c)^2, \label{eq:UV_Gamma2_2}
\end{align}
with the curvature masses defined in Eqs.~(\ref{eq:curvature_masses})-(\ref{eq:curvature_masses_2}). The flow equations for the two-point functions are then solved using the grid method, which, however, reduces to solving the flow equations only at a single grid point, i.e.~the IR minimum $\rho_0$, since they do not couple different grid points in the LPA setup.

Finally, the spectral functions are obtained from Eq.~(\ref{eq:spectral_function_2}) using the real-time two-point functions as obtained numerically within the LPA setup and as given analytically in the SC1L setup.

\section{Effective potential, couplings, and masses}
\label{sec:pot_couplings_masses}

In this section we present our results for the effective potential, the three-point coupling constants, as well as for the curvature masses and the pole masses. We begin with a discussion of the effective potential $U_k(\rho)$ for both the LPA setup as well as the SC1L setup, see Fig.~\ref{fig:pot_comparison}. In the UV, the effective potential already shows spontaneous chiral symmetry breaking as its global minimum is at large values of the $\sigma$ field. This is of course done by construction and necessary since the bosonic fluctuations tend to restore chiral symmetry. Each of the two UV potentials is chosen so as to obtain the same values for the pion decay constant and the pole masses in the IR. We also note that the UV potential in the SC1L setup starts out less strongly broken than in the LPA setup. This can be understood by considering the different propagators that enter the flow equations for the effective potential: In the SC1L setup, the propagators contain the one-loop self-energies which tend to suppress the bosonic fluctuations and hence the resulting symmetry restoration.
 
\begin{figure}[t]
	\includegraphics[width=0.48\textwidth]{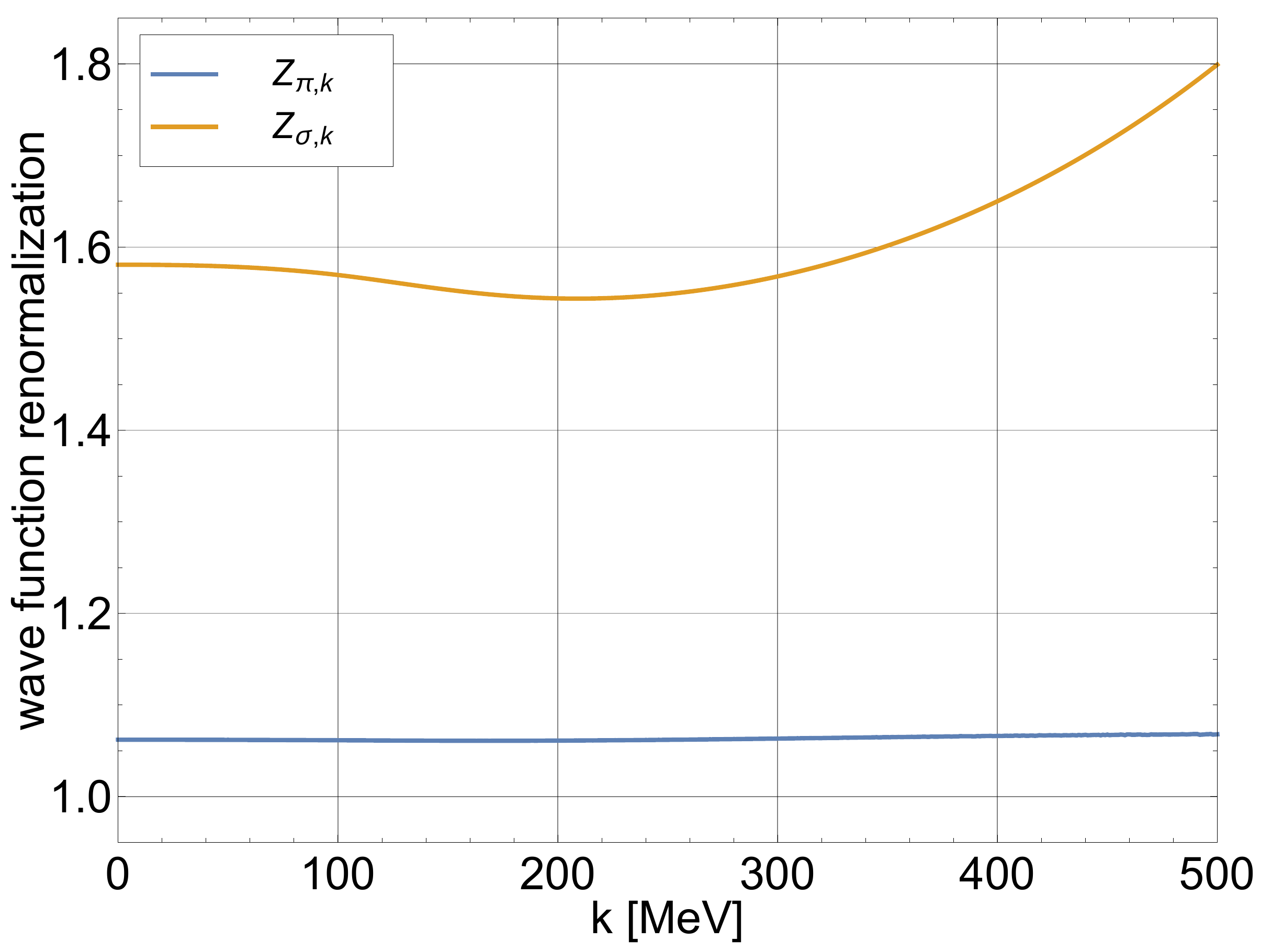}
	\caption{The wave function renormalization factors $Z_{\pi,k}$ and $Z_{\sigma,k}$ are shown as a function of the RG scale $k$ for the self-consistent one-loop setup.}
	\label{fig:Z_k}
\end{figure}

\begin{figure}[t]
	\includegraphics[width=0.48\textwidth]{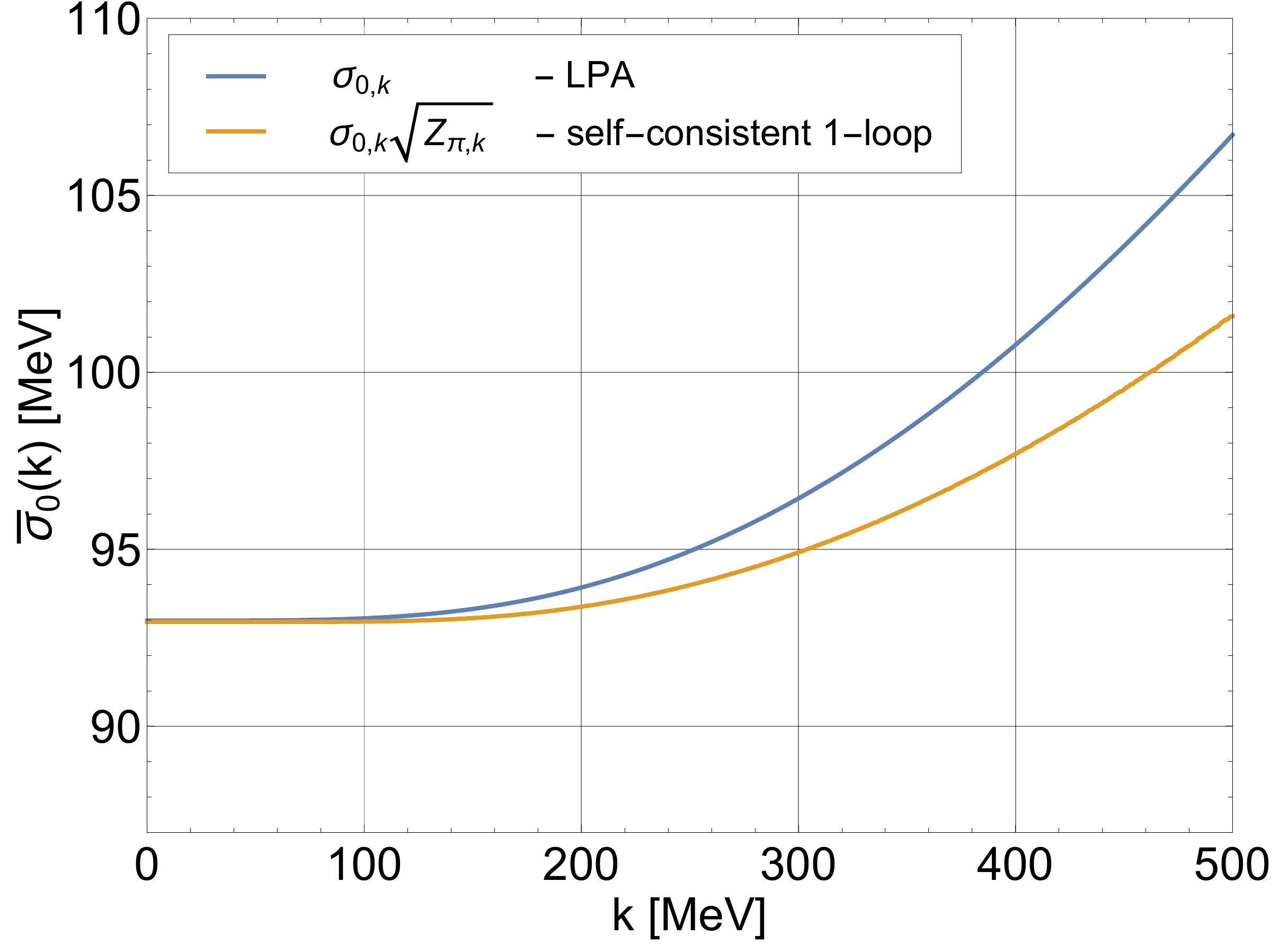}
	\caption{The scale-dependent minimum of the effective potential for the LPA setup, $\sigma_{0,k}$, is shown together with the renormalized minimum obtained for the self-consistent one-loop setup $\bar{\sigma}_{0,k}$.}
	\label{fig:minimum}
\end{figure}

From the effective potential we obtain the pion decay constant $f_\pi$ which is identified with the value $\sigma_0$ of the $\sigma$ field at its global minimum. In the case of the SC1L setup, this value has to be renormalized due to the appearance of non-trivial wave function renormalization factors $Z_\pi$ and $Z_\sigma$. These factors arise from the non-trivial momentum dependence of the SC1L propagators and are defined as
\begin{align}
\label{eq:Zs}
Z_{\pi,k}&=\left.\frac{\partial \Gamma^{(2)}_{\pi,k}(p^2)}{\partial p^2}\right|_{p=0},\\
Z_{\sigma,k}&=\left.\frac{\partial \Gamma^{(2)}_{\sigma,k}(p^2)}{\partial p^2}\right|_{p=0}.
\end{align}
The renormalized minimum of the effective potential can then be obtained using the pion wave function renormalization factor,
\begin{align}
\label{eq:renormalized_minimum}
\bar{\sigma}_{0,k}\equiv \sigma_{0,k}\sqrt{Z_{\pi,k}}.
\end{align}
While in the LPA setup we have $Z_{\pi,k}=Z_{\sigma,k}=1$ by definition, the wave function renormalization factors for the SC1L setup are shown in Fig.~\ref{fig:Z_k} as a function of the RG scale $k$. We note that $Z_{\pi,k}$ is almost scale-independent and close to 1 and therefore only has a small effect on the renormalization of the sigma field. The sigma wave function renormalization on the other hand shows a non-trivial scale dependence and is significantly larger, with an IR value of $Z_{\sigma,k=0}=1.58$ as opposed to $Z_{\pi,k=0}=1.06$ for the pion.

\begin{figure}[t]
	\includegraphics[width=0.48\textwidth]{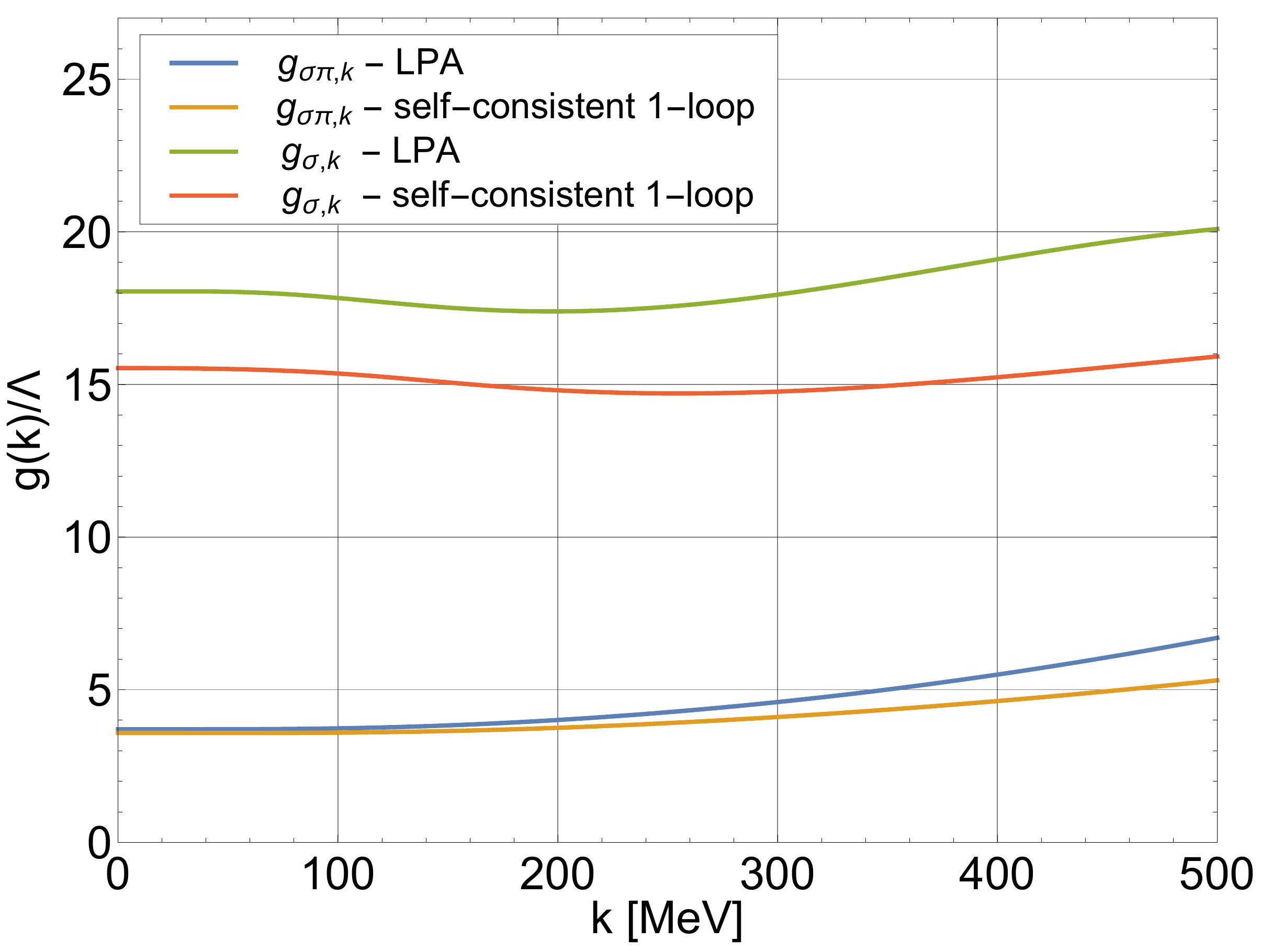}
	\caption{Flow of the three-point coupling constants $g_{\sigma\pi\pi,k}$ and 
	 $g_{\sigma\sigma\sigma,k}$ for the LPA and the self-consistent one-loop setup.}
	\label{fig:couplings}
\end{figure}

\begin{figure*}[t]
	\includegraphics[width=0.48\textwidth]{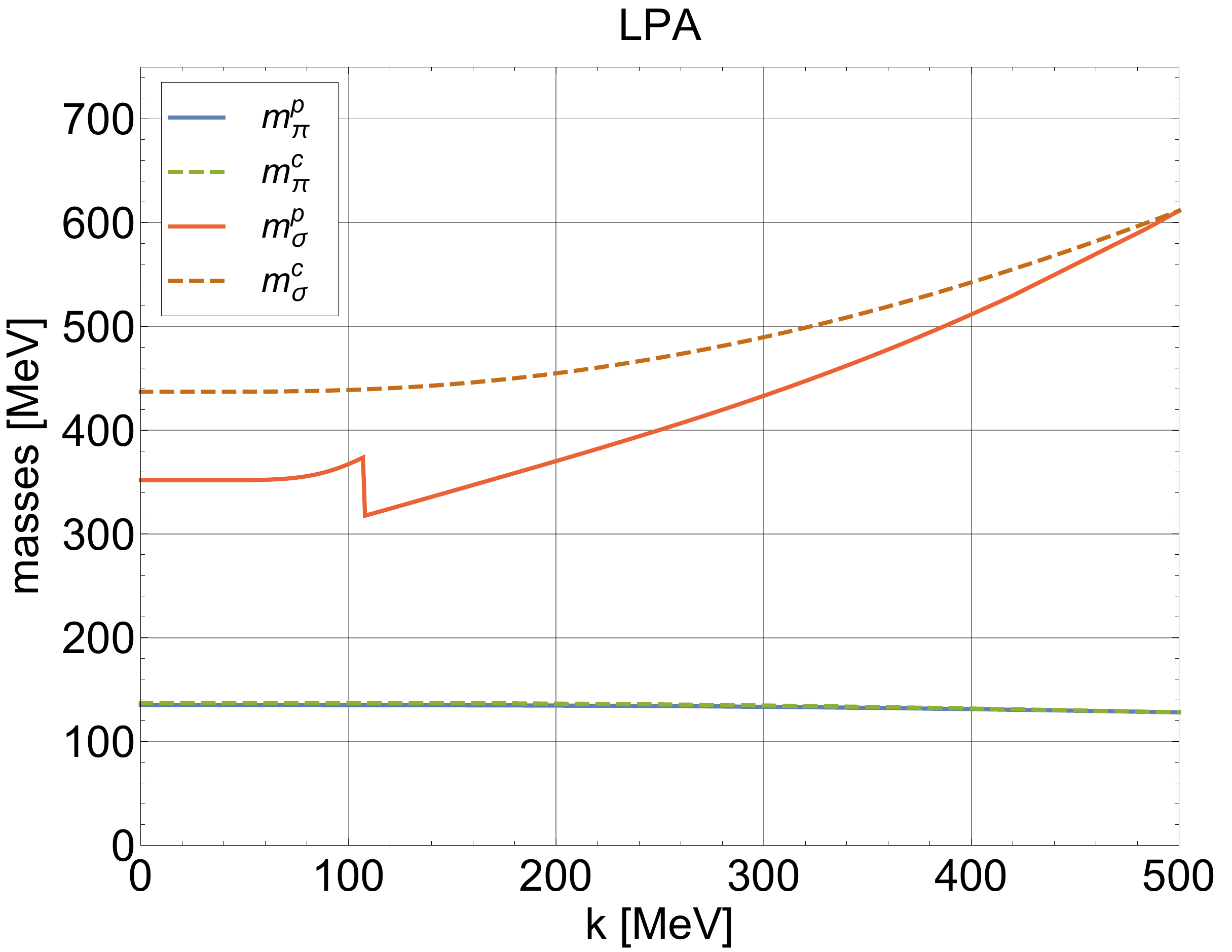}
	\includegraphics[width=0.48\textwidth]{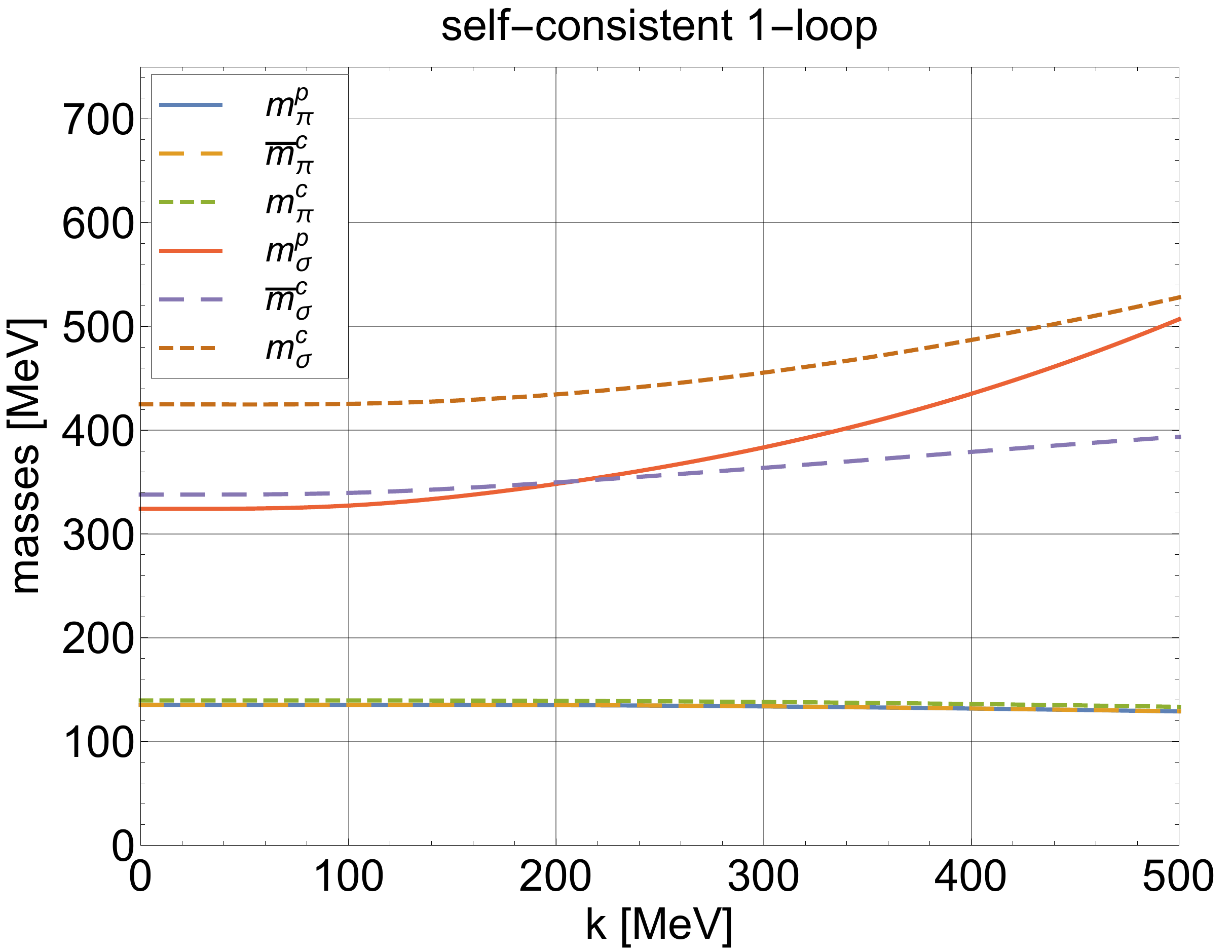}
	\caption{The various masses obtained within the LPA (left) and the self-consistent one-loop setup (right) are shown vs.~the RG scale $k$. For the LPA setup we show the Euclidean curvature masses in comparison to the real-time pole masses while for the SC1L setup we also show the renormalized curvature masses which involve the wave function renormalization factors, see text for details.}
	\label{fig:masses}
\end{figure*}

Given the wave function renormalization factors, we can compute the renormalized minimum $\bar{\sigma}_{0,k}$ of the effective potential and compare it to the LPA result $\sigma_{0,k}$, see Fig.~\ref{fig:minimum}. As also evident from Fig.~\ref{fig:pot_comparison}, the location of the minimum flows from larger values in the UV to smaller values in the IR, with the LPA minimum starting at larger values than the SC1L minimum. In the IR, the LPA minimum $\sigma_{0,k}$ and the renormalized SC1L minimum $\bar{\sigma}_{0,k}$ arrive at the same value of $f_\pi= 93$~MeV since the respective parameter sets have been chosen accordingly.

From the scale-dependent effective potential $U_k(\rho)$, we can also determine the coupling constants $g_{\sigma,k}$ and $g_{\sigma\pi,k}$, cf.~(\ref{eq:couplings}) with (\ref{eq:couplings-s})-(\ref{eq:couplings-sp}). Their scale dependence is shown in Fig.~\ref{fig:couplings} for the LPA and the SC1L setup. We note that $g_{\sigma,k}$ is about three times larger than $g_{\sigma\pi,k}$ and that the  SC1L couplings tend to be smaller than the LPA couplings, see Tab.~\ref{tab:IR_values} for explicit values.

\begin{table}[b]
	\centering
	\begin{tabular}{C{0.8cm}|C{0.65cm}|C{0.65cm}|C{0.85cm}|C{0.85cm}|C{0.85cm}|C{0.85cm}|C{0.85cm}|C{0.85cm}}
		& $g_{\sigma}$ & $g_{\sigma\pi}$ & $m_{\pi}^c$  & $\bar{m}_{\pi}^c$ & $m_{\pi}^p$ & $m_{\sigma}^c$ & $\bar{m}_{\sigma}^c$  & $m_{\sigma}^p$    \\
		& [$\Lambda$] & [$\Lambda$] & [MeV]& [MeV] & [MeV] & [MeV] & [MeV] & [MeV]
		\\
		\hline\hline
		LPA &  18.0  & 3.70 &  137 & - & 135 & 437 & - & 352  \\
		SC1L & 15.5  & 3.58 &  140 & 135 & 135 & 425 & 338 & 324
	\end{tabular}
	\caption{Numerical values for the coupling constants, the curvature masses, the renormalized curvature masses, and the pole masses as obtained with the LPA as well as with the self-consistent one-loop (SC1L) setup in the IR.}
	\label{tab:IR_values} 
\end{table}

We close this subsection with a discussion of the scale dependence of the various masses obtained within the LPA as well as within the SC1L setup, see Fig.~\ref{fig:masses}. 

\begin{figure*}[t]
	\includegraphics[width=0.48\textwidth]{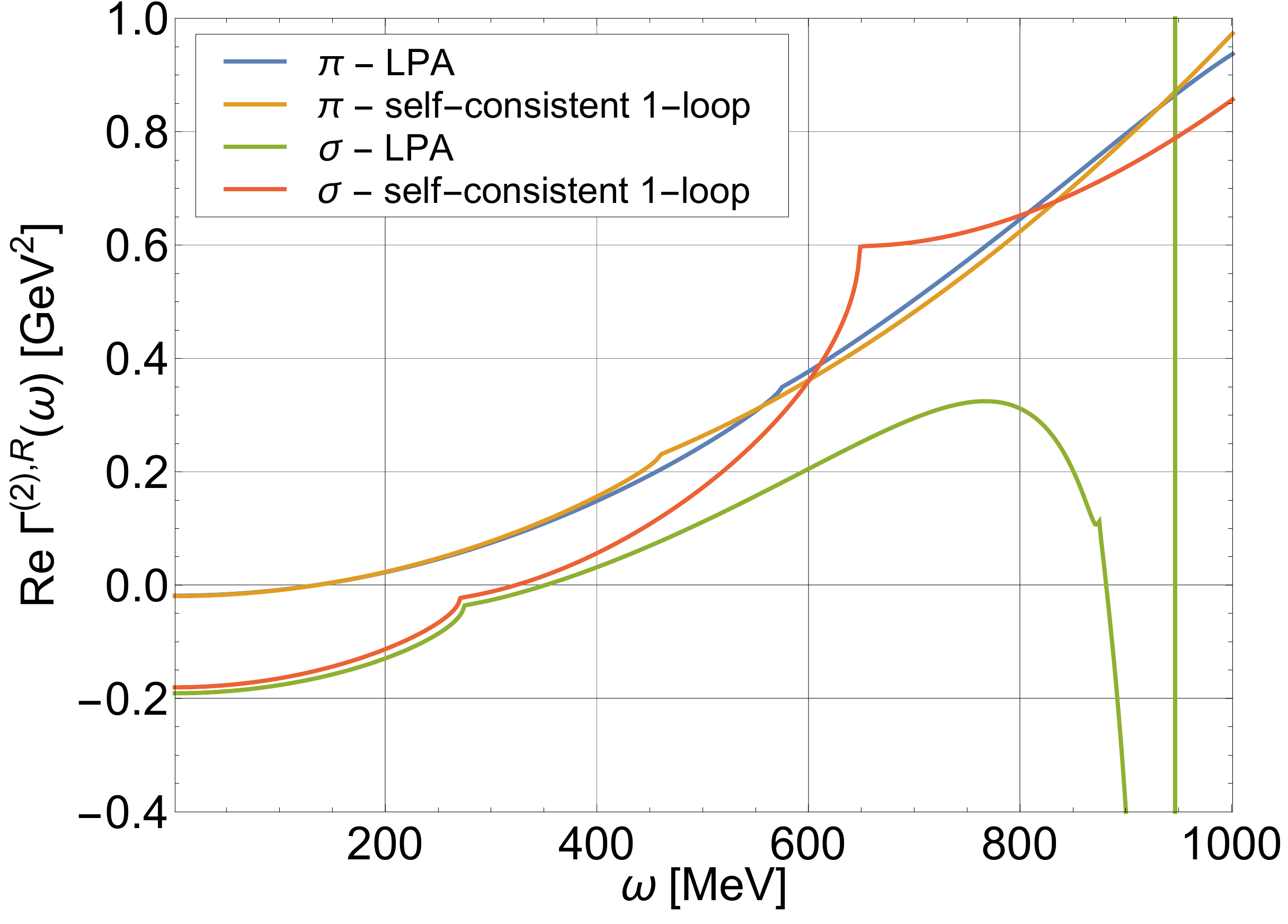}
	\includegraphics[width=0.48\textwidth]{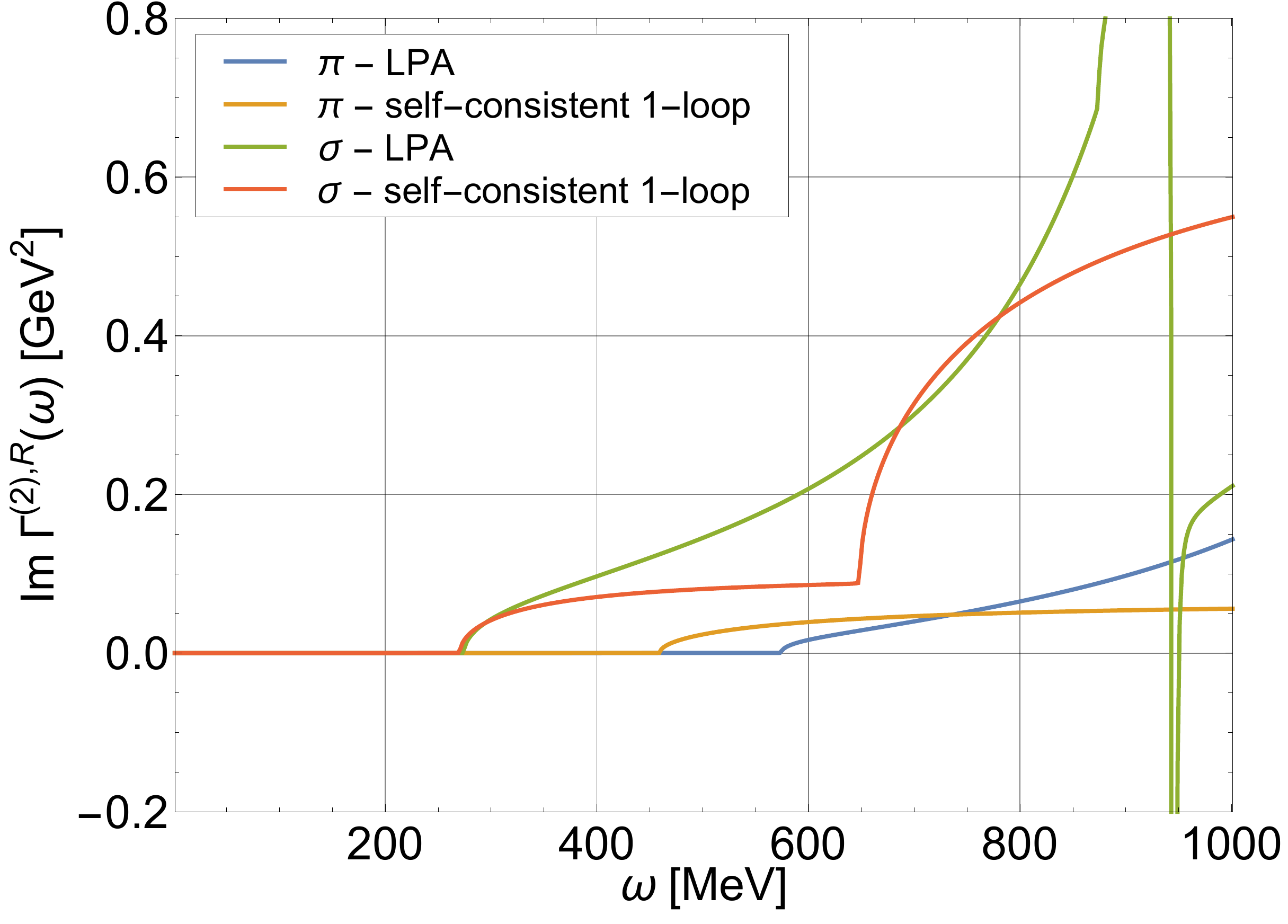}
	\caption{The real (left) and the imaginary (right) part of the pion and sigma two-point functions are shown vs.~external energy $\omega$ in the IR as obtained from the LPA setup as well as from the self-consistent one-loop setup.}
	\label{fig:Gamma2}
\end{figure*}

For the LPA setup we show the Euclidean curvature masses as obtained from the effective potential and the real-time pole masses as obtained from the zero crossings of the real parts of the corresponding two-point functions. In the UV, the LPA curvature and pole masses agree by construction. In the IR, the pole mass of the sigma meson is significantly smaller than its curvature mass while the pion pole mass stays close to its curvature mass, see also Tab.~\ref{tab:IR_values}. We note that the sigma meson pole mass suddenly jumps to a higher value at $k\approx 110$~MeV which can be understood as follows. In the UV, the sigma meson starts out as a stable particle in the LPA setup with its spectral function given by a Dirac-delta function. This delta function then moves to smaller energies, i.e.~pole masses, as the RG scale $k$ is decreased. Eventually, the two-pion decay threshold, which also moves to smaller energies due to the decreasing pion curvature mass, overtakes the delta peak which leads to the formation of a broad resonance structure in the spectral function. This also leads to a sudden increase of the pole mass as visible in Fig.~\ref{fig:masses}.

\begin{figure*}[t]
	\includegraphics[width=0.48\textwidth]{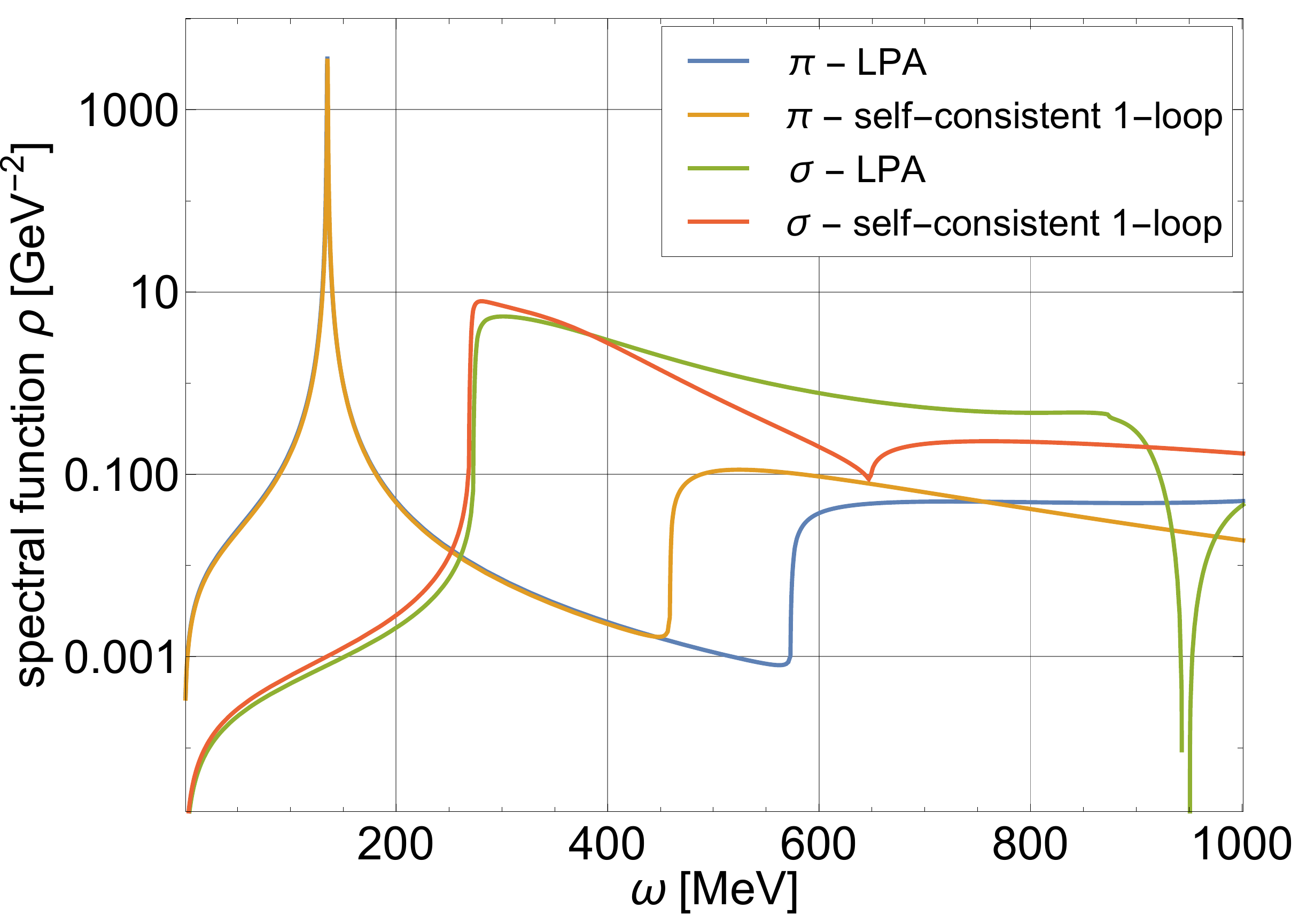}
	\includegraphics[width=0.48\textwidth]{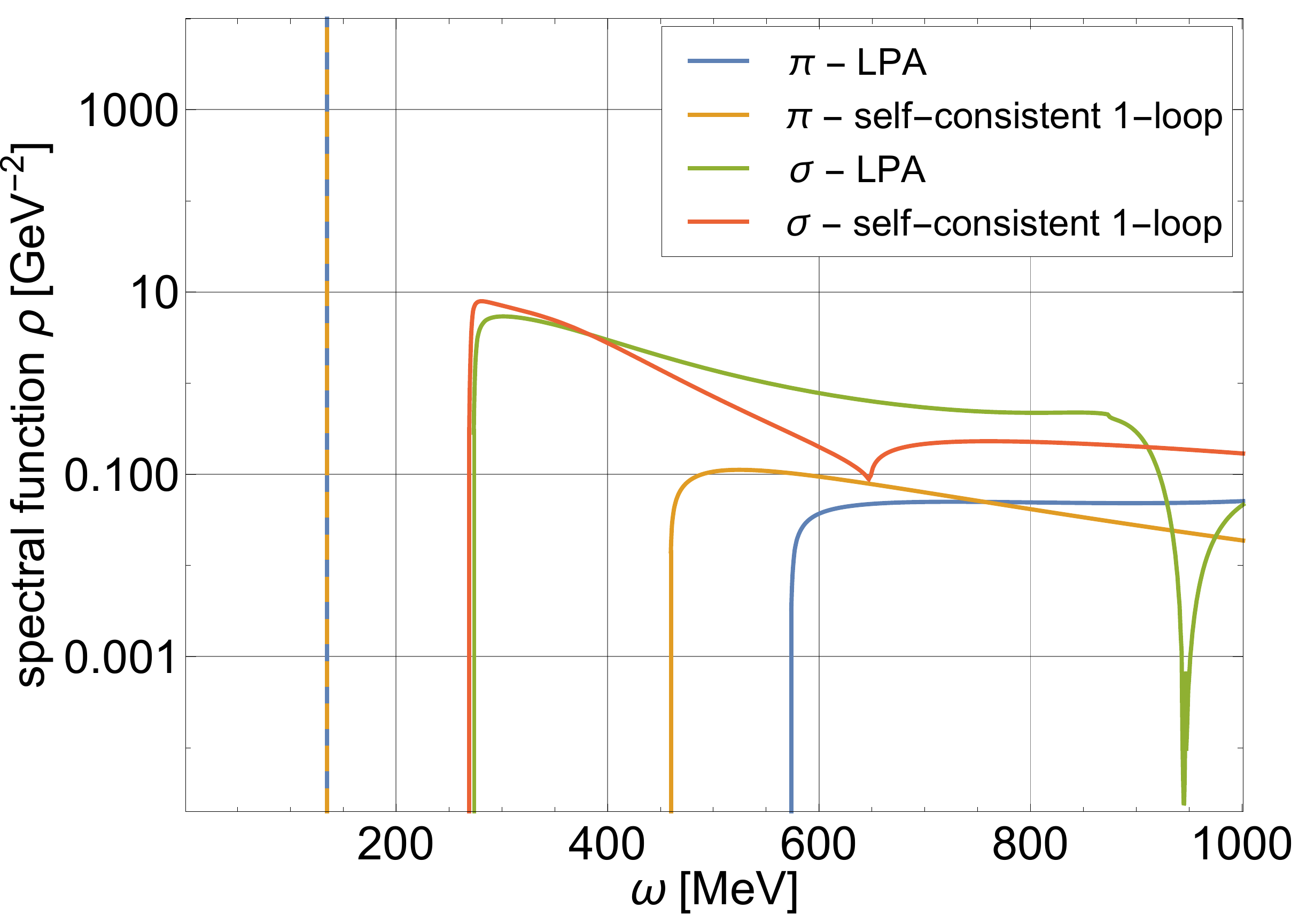}
	\caption{The pion and sigma spectral functions are shown vs.~external energy $\omega$ in the IR as obtained from the LPA setup as well as from the self-consistent one-loop setup for $\varepsilon=0.1$~MeV (left) and for $\varepsilon=0$~MeV (right), see text for details.}
	\label{fig:spectral}
\end{figure*}

On the right-hand side of Fig.~\ref{fig:masses} we show the corresponding masses obtained within the SC1L setup. In this case we also show the renormalized curvature masses which are defined as
\begin{align}
\label{eq:renormalized_curvature_masses}
\bar{m}^{c}_{\pi,k}\equiv m^{c}_{\pi,k}/\sqrt{Z_{\pi,k}},\\
\bar{m}^{c}_{\sigma,k}\equiv m^{c}_{\sigma,k}/\sqrt{Z_{\sigma,k}},
\end{align}
an thus take the effect of the wave function renormalization factors into account. We note that, already in the UV, the masses are different from one another since the UV two-point functions, which determine the pole masses and the wave function renormalizations, already include the one-loop self energies and thus go beyond LPA. As in the LPA setup, however, the pion masses remain almost constant throughout the flow. The sigma meson masses show a stronger scale dependence and continuously decrease towards the IR, see Tab.~\ref{tab:IR_values} for explicit values.
\section{Two-point functions and spectral functions}
\label{sec:2PF_and_SF}
We now turn to our results for the real-time two-point functions and spectral functions. In Fig.~\ref{fig:Gamma2} we show the real and the imaginary part of the sigma and pion two-point functions in the IR as obtained from the LPA aFRG flow as well as the SC1L setup. While the two frameworks give similar results at low external energies $\omega$, one encounters undesired effects from the UV cutoff in the LPA two-point functions at larger energies. For the SC1L setup, on the other hand, the two-point functions are well defined for arbitrarily large (and small) energies. We also note that the two-point functions show several kinks which correspond to the location of decay thresholds in the spectral functions. These decay thresholds are determined by the curvature masses in the LPA and by the pole masses in the SC1L setup, respectively. They are more clearly visible in the imaginary parts of the two-point functions as shown on the right-hand side of Fig.~\ref{fig:Gamma2}.

\begin{figure*}[t]
	\includegraphics[width=0.49\textwidth]{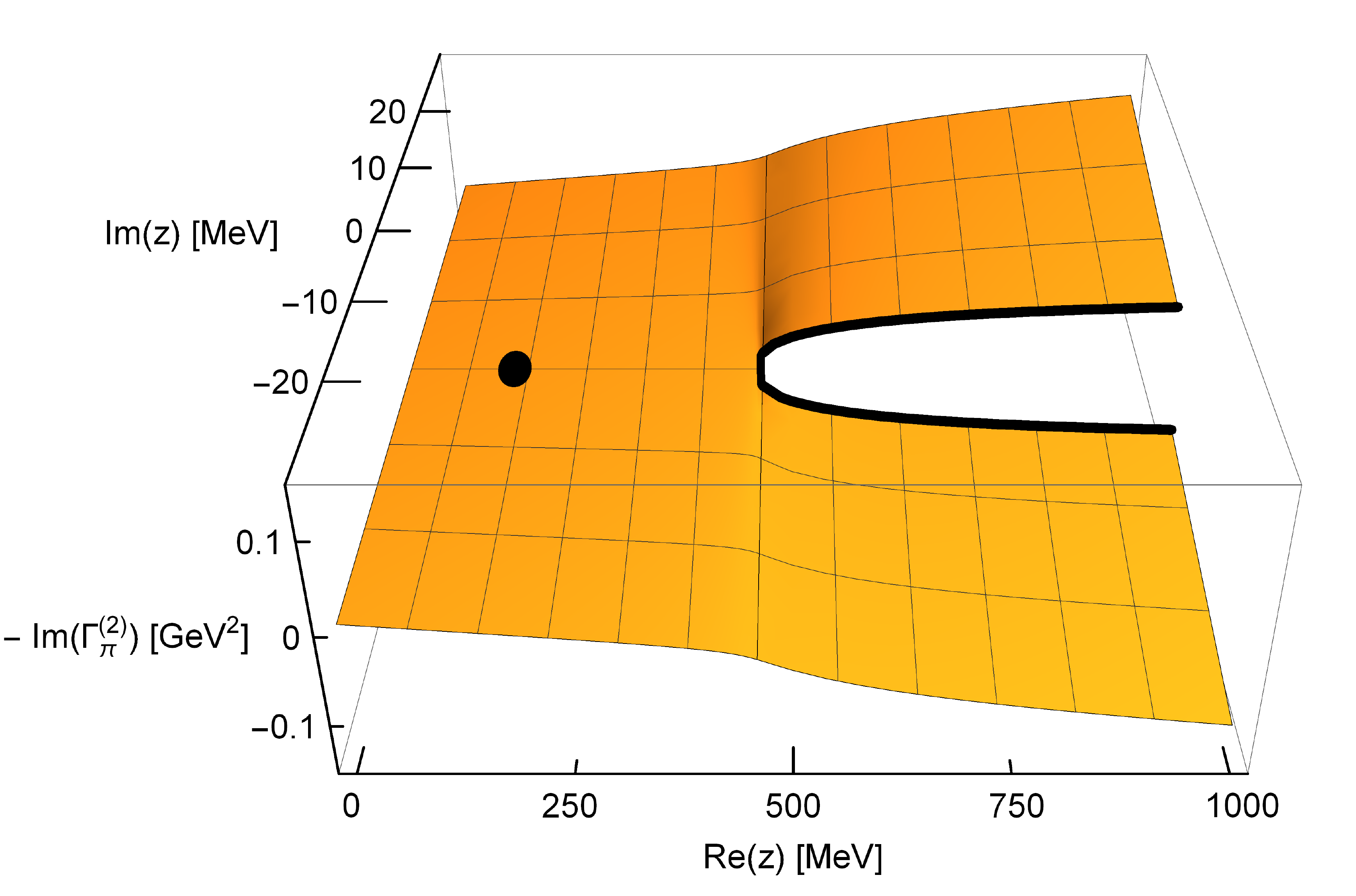}
	\includegraphics[width=0.49\textwidth]{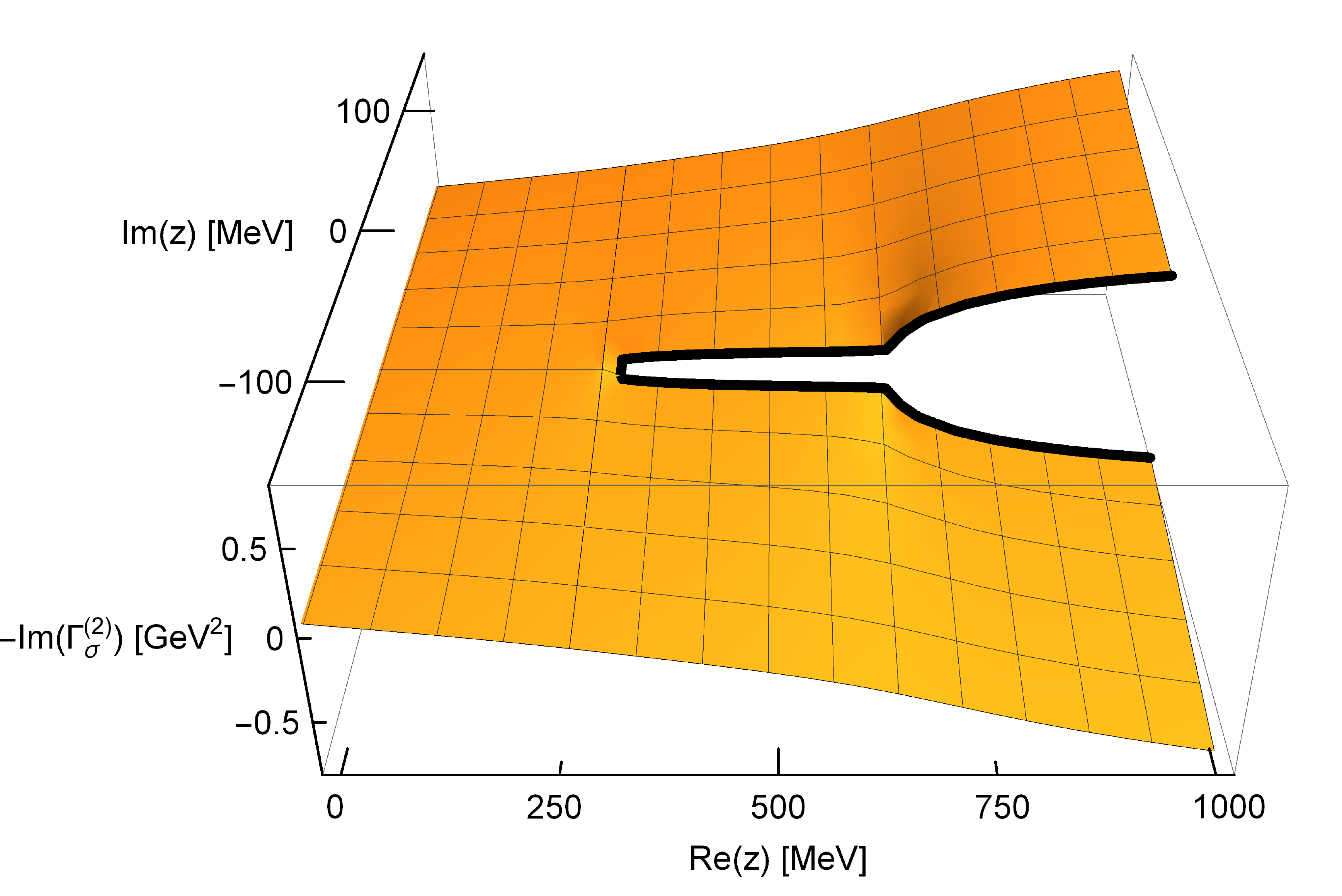}
	\caption{The imaginary part of the pion (left) and the sigma (right) two-point function is shown in the complex momentum plane. The pion two-point function exhibits a branch cut corresponding to the $\pi^*\rightarrow \sigma+\pi$ decay process while the sigma two-point function exhibits two branch cuts corresponding to the processes $\sigma^*\rightarrow \pi+\pi$ and $\sigma^*\rightarrow \sigma+\sigma$. The complex pole of the pion is marked as a black dot.}
	\label{fig:pion_Riemann_1}
\end{figure*}

The first threshold in the sigma two-point function is determined by the process 
\begin{align}
\label{eq:process_1}
\sigma^*\rightarrow \pi+\pi, \qquad \omega \geq 2 m_\pi,
\end{align}
i.e.~the decay of an off-shell sigma meson with energy $\omega$ into two on-shell pions which is kinematically possible when $\omega$ is larger than two times the pion mass. The second threshold in the sigma two-point function corresponds to the process 
\begin{align}
\label{eq:process_2}
\sigma^*\rightarrow \sigma+\sigma, \qquad \omega \geq 2 m_\sigma,
\end{align}
i.e.~the decay of an off-shell sigma meson into two on-shell sigma mesons. The pion two-point function, on the other hand, is only affected by a single process, i.e.~the decay of an off-shell pion into a sigma meson and a pion,
\begin{align}
\label{eq:process_3}
\pi^*\rightarrow \sigma+\pi, \qquad \omega \geq m_\sigma + m_\pi.
\end{align}

We now turn to the resulting spectral functions which are shown in Fig.~\ref{fig:spectral}. On the left-hand side we show the spectral functions  obtained from the LPA aFRG flow and SC1L flow for a small but finite value of $\varepsilon=0.1$~MeV. Both approaches again give very similar results at lower energies while the LPA approach breaks down at energies close to the UV cutoff scale. The pion spectral function shows a sharp peak at the pion pole mass followed by the $\pi^*\rightarrow \sigma+\pi$ decay threshold and the corresponding continuum part. Note that this decay threshold occurs at different energies in the two different calculations. With the analytically continued LPA  flows it is always determined by the curvature-mass parameters, and it therefore here occurs at $\omega= m_\sigma^c + m_\pi^c \approx 575$~MeV. With the self-consistent one-loop flows of the SC1L setup, on the other hand, it is determined by (the real parts of) their pole masses and hence occurs at $\omega = m_\sigma^p + m_\pi^p\approx 460$~MeV. This most significant difference is a direct consequence of the self-consistency of the SC1L flows for the two-point functions in the timelike domain. 

In contrast, less noticeable differences are observed in the sigma spectral function. Because the difference between curvature mass defined at $p^2=0$ and pole mass at $-p^2 = (m_\pi^p)^2 $ is negligible for the pion in our calculations, as expected for the lightest meson, the $\sigma^*\rightarrow \pi+\pi$ thresholds occur at practically the same energies in both calculations. The small differences in the sigma-meson spectral function near this two-pion threshold arise from the slightly different resonance pole positions of the sigma meson in both calculations, see Table \ref{tab:IR_values}. Only visible in the self-consistent SC1L calculation, on the other hand, is the $\sigma^*\rightarrow \sigma+\sigma$ threshold at higher energies where the analytically continued LPA flow starts to become cutoff sensitive and less reliable.  

In order to exemplify the influence of the small but finite $\varepsilon >0$ in these results, on the right-hand side of Fig.~\ref{fig:spectral} we also show the corresponding results obtained in the limit $\varepsilon\rightarrow 0$. In the case of the LPA aFRG flows, this result was obtained by performing the limit $\varepsilon\rightarrow 0$ analytically in the flow equations for the imaginary parts of the two-point functions, see for example \cite{Jung:2016yxl} for details. In the case of the SC1L setup this limit can readily be obtained numerically since the two-point functions are known analytically. In contrast to the $\varepsilon>0$ results on the left, the pion peak is now given by a Dirac delta function and the spectral functions otherwise vanish below the lowest-lying decay thresholds. At energies larger than these lowest-lying thresholds, however, the spectral functions are basically unaffected by this limit. We note that, at finite temperature, a finite value for $\varepsilon$ may be used to qualitatively mimic the effects from multi-particle scattering processes in the thermal medium.

\begin{figure*}[t]
	\includegraphics[width=0.32\textwidth]{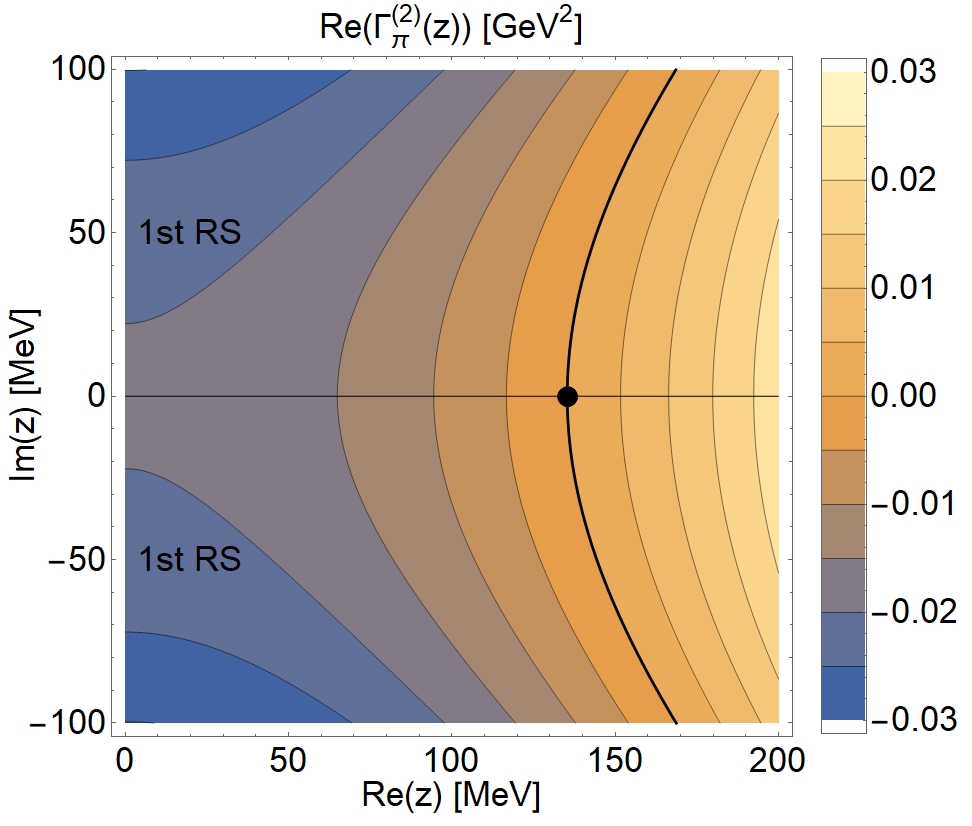}
	\includegraphics[width=0.32\textwidth]{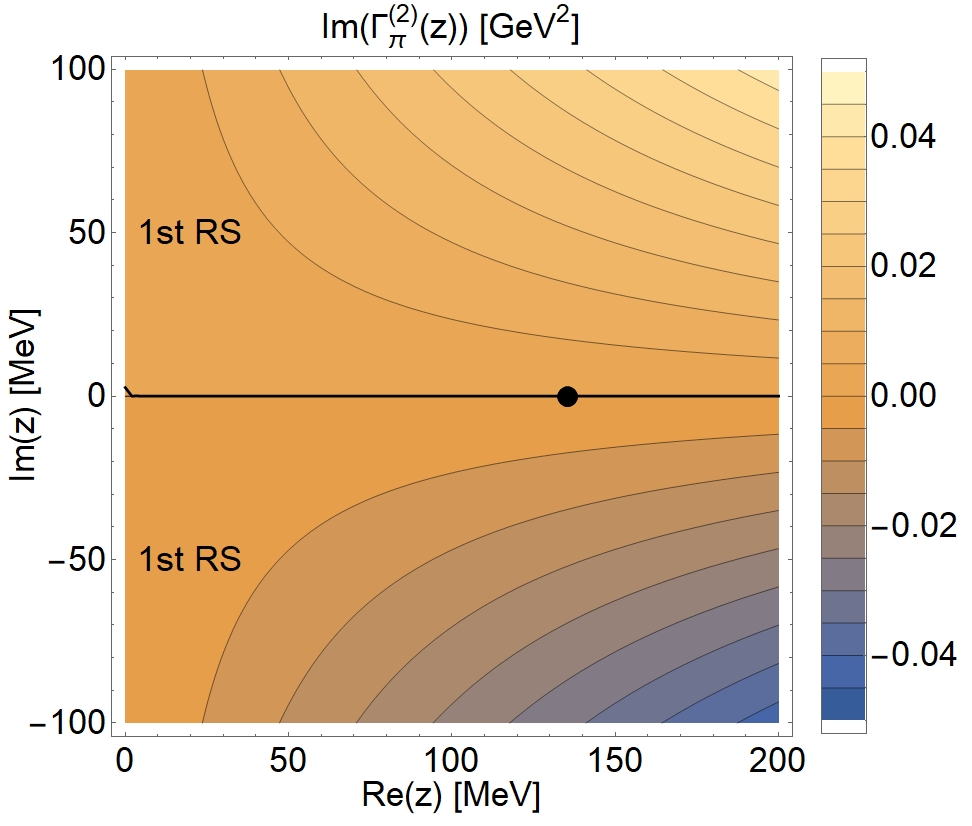}
	\includegraphics[width=0.32\textwidth]{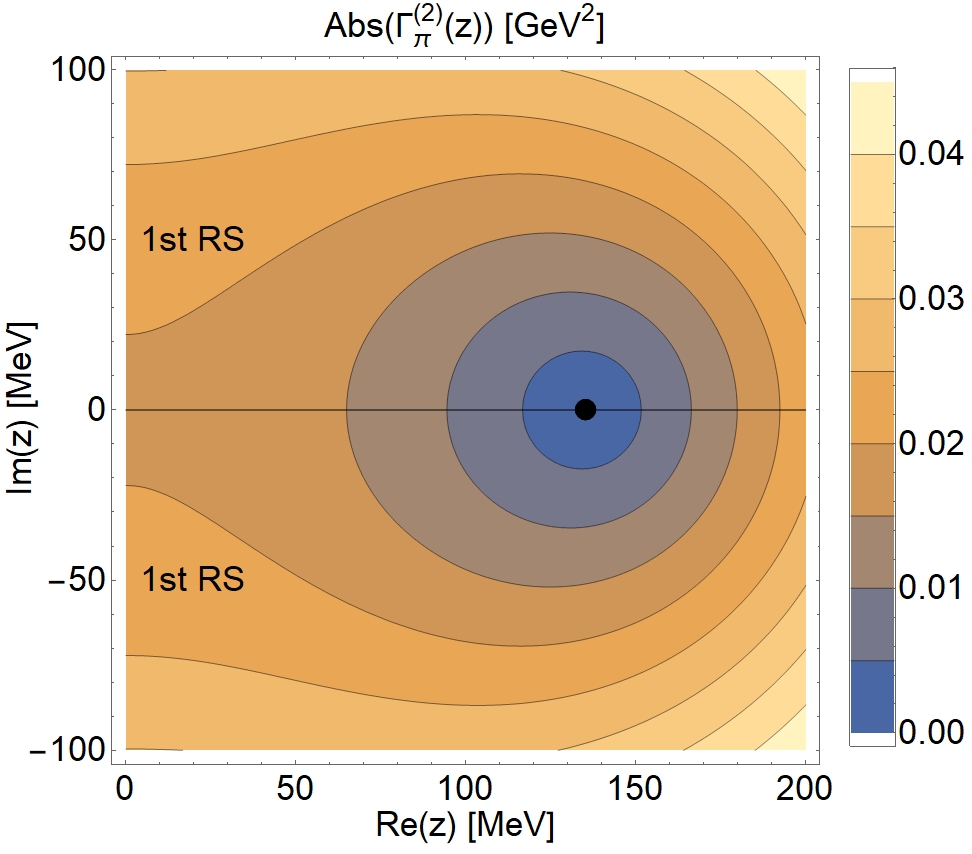}
	\caption{The real part (left), the imaginary part (middle), and the absolute value (right) of the pion two-point function are shown as a function of complex momentum $z$ as a contour plot on the first Riemann sheet (1st RS), see text for details.}
	\label{fig:pion_complex_plane}
\end{figure*}

\begin{figure*}[t]
	\includegraphics[width=0.32\textwidth]{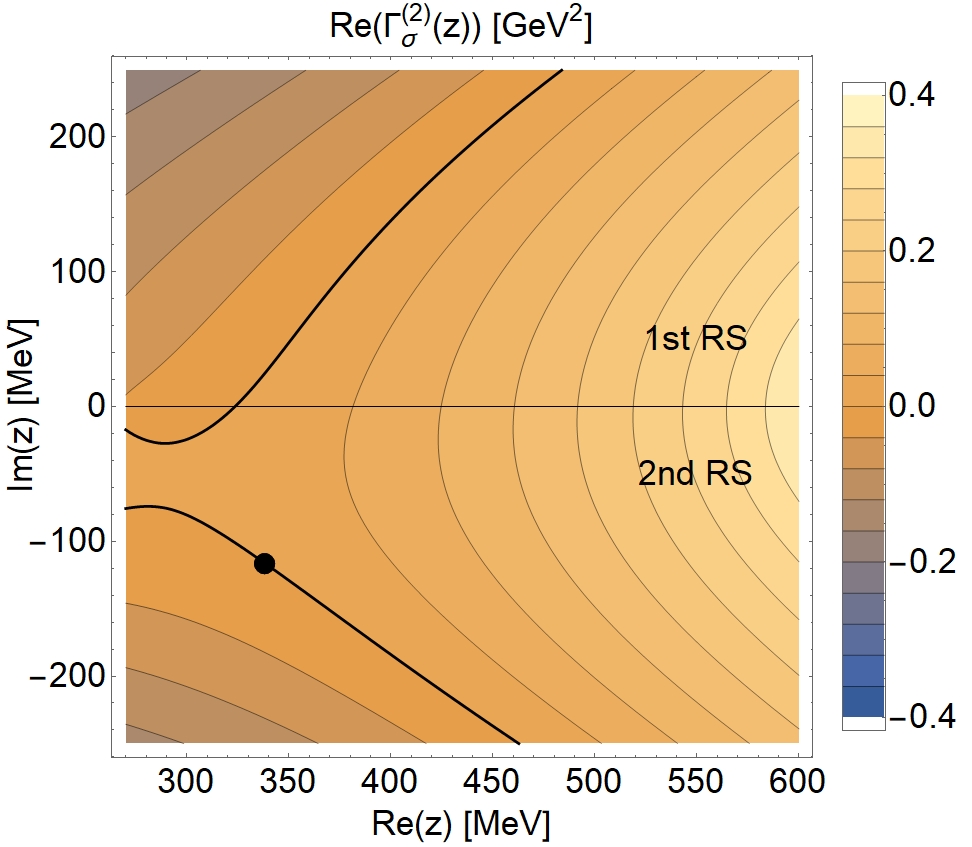}
	\includegraphics[width=0.32\textwidth]{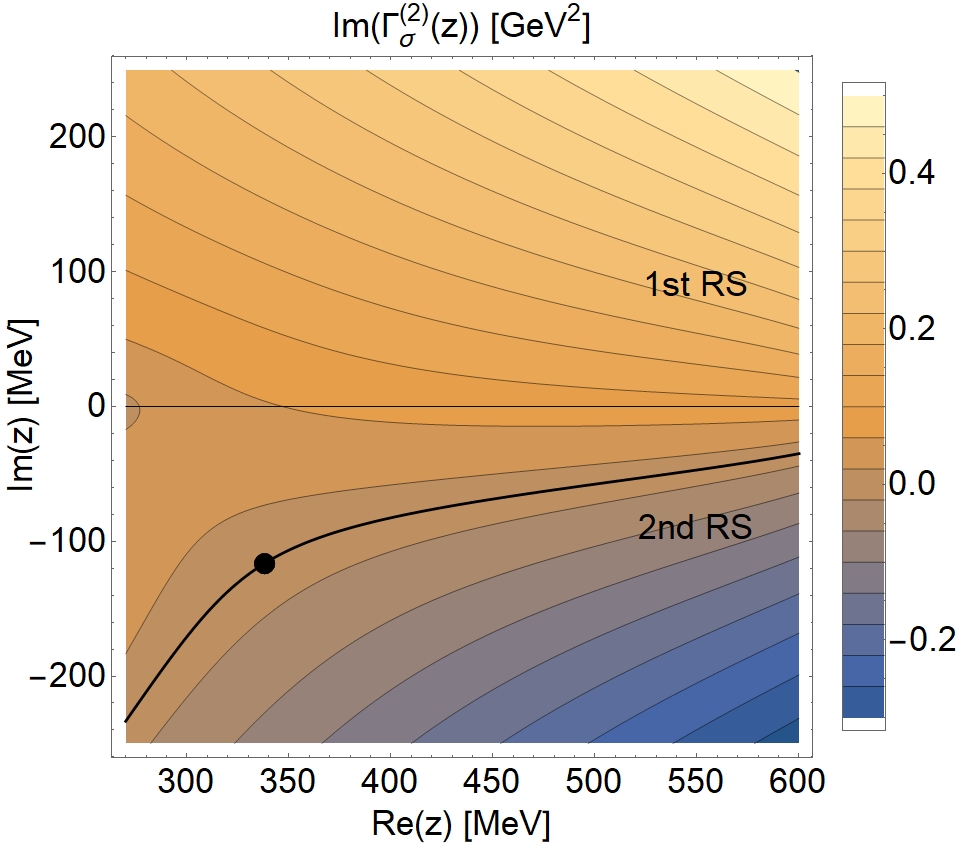}
	\includegraphics[width=0.32\textwidth]{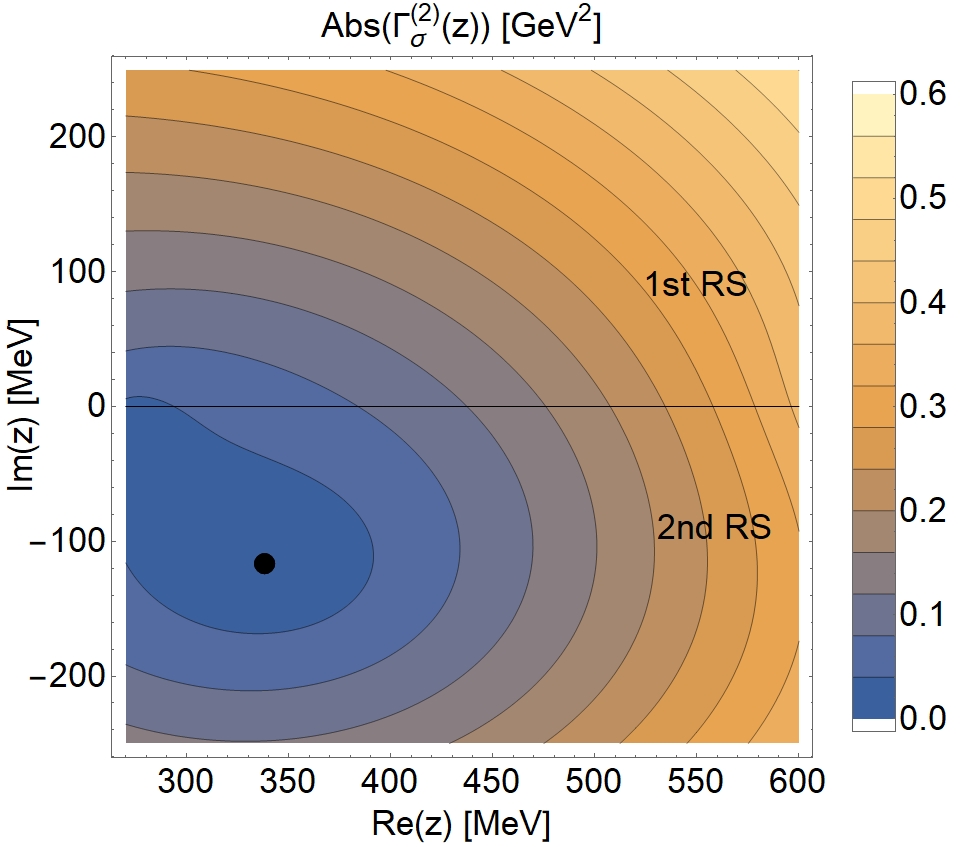}
	\caption{The real part (left), the imaginary part (middle), and the absolute value (right) of the sigma two-point function are shown as a function of complex momentum $z$ as a contour plot. The upper half-plane shows the first Riemann sheet (1st RS) while the lower half-plane shows the second Riemann sheet (2nd RS), see text for details.}
	\label{fig:sigma_complex_plane}
\end{figure*}

\section{Analytic structure in the complex plane}
\label{sec:analytic_structure}
In this section we discuss the analytic structure of the real-time sigma and pion two-point functions in the complex plane of the analytically continued energy variable $z$, with $z=\omega $ along the timelike real axis, as obtained from the self-consistent one-loop aFRG setup. Analyzing this structure is possible since these functions are given in analytic form, cf.~Eqs.~(\ref{eq:1L_2PF})-(\ref{eq:1L_2PF_2}), while they are only known numerically in the LPA setup. We begin by studying the branch-cut structure of the imaginary part of the two-point functions which is shown in Fig.~\ref{fig:pion_Riemann_1}. The pion two-point function shows a branch cut starting at $\omega=\text{Re}\:z=m_{\pi,k=0}^{p}+m_{\sigma,k=0}^{p}\approx 460$~MeV corresponding to the process $\pi^*\rightarrow \sigma+\pi$ while the sigma two-point functions shows two branch cuts, one starting at $\omega=\text{Re}\:z=2m_{\pi,k=0}^{p}\approx 270$~MeV corresponding to the process $\sigma^*\rightarrow \pi+\pi$ and one at $\omega=\text{Re}\:z=2m_{\sigma,k=0}^{p}\approx 650$~MeV corresponding to the process $\sigma^*\rightarrow \sigma+\sigma$. The pion pole is represented by a black dot located on the real axis at $\omega=m^p_{\pi,k=0}=135$~MeV. The complex pole of the sigma meson is not visible since it is not located on the so-called first Riemann sheet which is shown here.

In Fig.~\ref{fig:pion_complex_plane} we show contour plots of the real part, the imaginary part, and the absolute value of the pion two-point function evaluated at complex $z$ as analytic continuation of the energy $\omega$ along the real axis. The energy range is chosen such as to encompass the pole of the pion propagator, i.e.~the zero-crossing of the two-point function. In this energy range no decay thresholds are present and hence only the first Riemann sheet is shown. We note that the real part is symmetric about the real axis and that the crossing of the zero-line with the real axis defines the pion pole. The imaginary part, on the other hand, is antisymmetric with respect to the real axis and the zero-line coincides with the real axis. Finally, the absolute value of the pion two-point function clearly shows the location of the pion pole as the point in the complex momentum plane where both the real part as well as the imaginary part vanishes.

\begin{figure}[t]
	\includegraphics[width=0.48\textwidth]{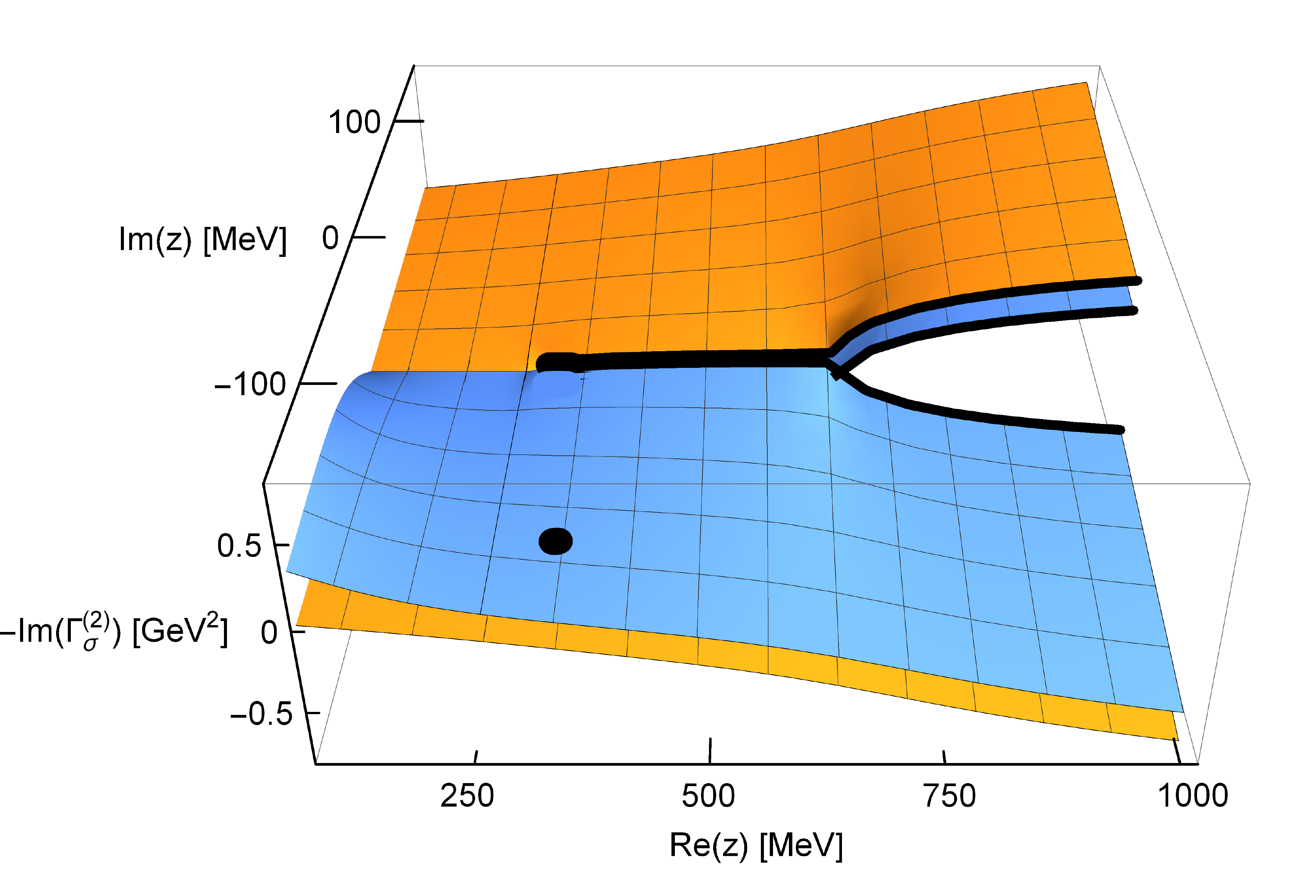}
	\caption{The first (orange) and the second (blue) Riemann sheet of the imaginary part of the sigma two-point function is shown in the complex momentum plane. The complex pole of the sigma meson is marked as a black dot on the second Riemann sheet, see text for details.}
	\label{fig:sigma_Riemann_1}
\end{figure}

We now turn to the sigma two-point function, which is shown in Fig.~\ref{fig:sigma_complex_plane}, again for complex energy $z$. This time, we show an energy range between the $\sigma^*\rightarrow \pi+\pi$ threshold and the $\sigma^*\rightarrow \sigma+\sigma$ where two Riemann sheets are of importance. Crossing continuously across the cut along the real axis from above one crosses from the first Riemann sheet in the upper half plane to the second Riemann sheet in the lower half plane, cf.~App.~\ref{app:one_loop} for a discussion of the different sheets. The complex pole of the sigma meson is then found on the second Riemann sheet at 
\begin{align}
\label{eq:sigma_pole}
z_\sigma^p\, \approx \, 338-i\: 116 \;\; \text{MeV}.
\end{align}
We note that the same value can also be obtained using numerical reconstruction techniques such as the Schlessinger point method (SPM), see \cite{Schlessinger1966, Tripolt:2016cya, Tripolt:2018xeo}. This method can be directly applied to numerical data on the spectral function and thus can in principle also be used for numerical results from FRG flows such as our LPA calculation here.

The branch cut structure of the sigma two-point function is more readily visible in a three-dimensional plot, see Fig.~\ref{fig:sigma_Riemann_1} where the first and the second Riemann sheet of the imaginary part of the sigma two-point function are shown together with the location of the complex pole of the sigma meson. The two Riemann sheets are continuously connected along the branch cut corresponding to the process $\sigma^*\rightarrow \pi+\pi$ while they are discontinuous along the second branch cut corresponding to the process $\sigma^*\rightarrow \sigma+\sigma$. We note that there exist additional (infinitely many) Riemann sheets which can be accessed as described in App.~\ref{app:one_loop}, but which are of no further relevance for our present study.

\begin{figure}[t]
	\includegraphics[width=0.45\textwidth]{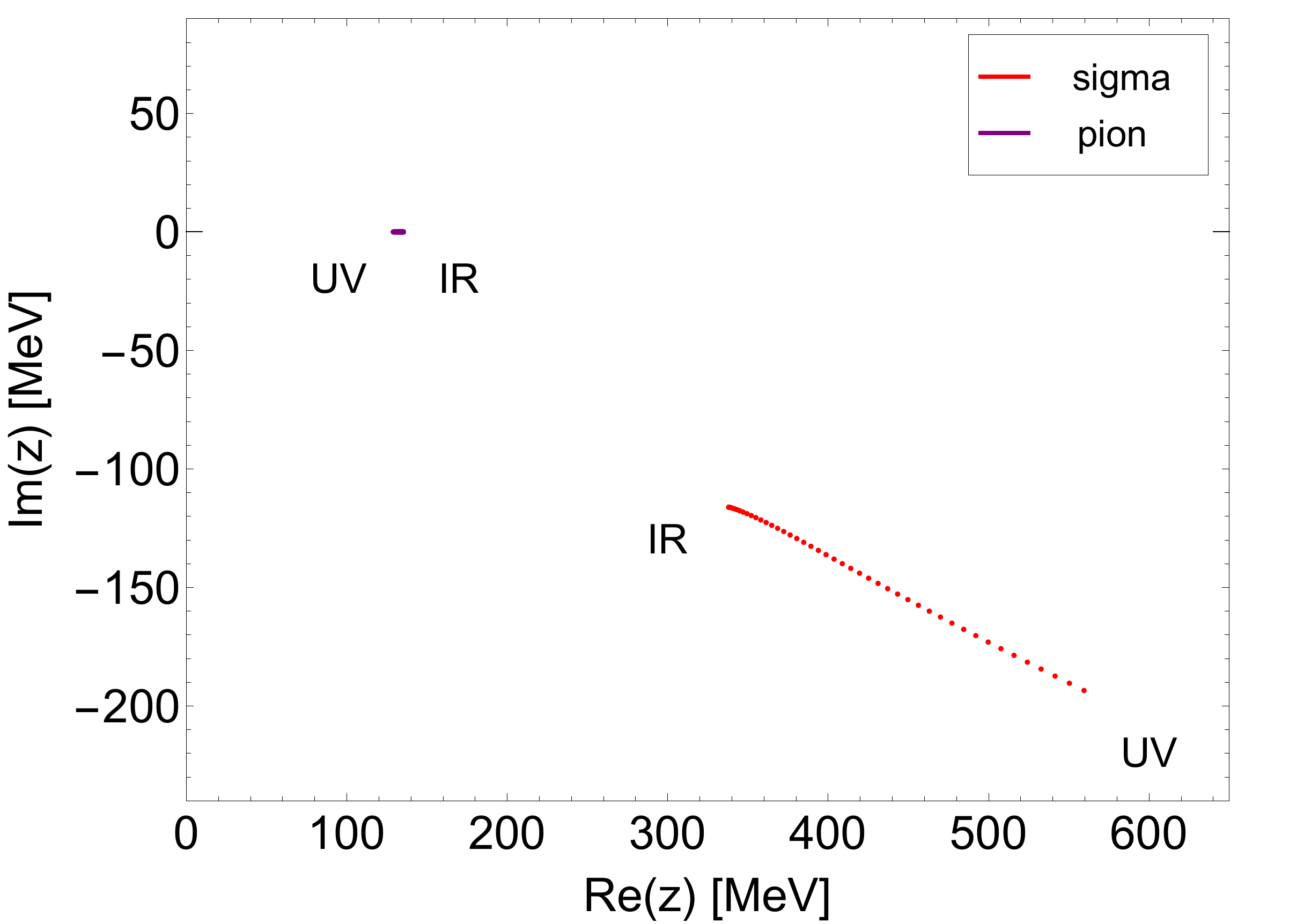}
	\caption{Flow of the pion pole and of the sigma meson resonance pole in the complex plane. The red dots mark equidistant steps of $\Delta k=10$~MeV in the FRG scale $k$, see text.}
	\label{fig:poles_k}
\end{figure}

We close this section by investigating the movement of the poles of the pion and the sigma meson in the complex plane as a function of the RG scale $k$, as shown in Fig.~\ref{fig:poles_k}. While the pion pole is always located on the real axis and moves from smaller values in the UV to slightly larger values in the IR, the complex pole of the sigma meson is located deep in the complex plane on the second Riemann sheet and moves from larger values in the UV to smaller values in the IR. We note that having access to the complete analytic structure of the two-point functions, including the location of the complex poles, represents a significant advantage over previous real-time FRG calculations and provides this  additional insight. In particular, the techniques developed in this work can be extended to phenomenologically more realistic effective theories, e.g.~involving additional (mesonic and baryonic) degrees of freedom at finite temperature and density \cite{Tripolt:2021jtp}.

\section{Towards fully self-consistent spectral functions}
\label{sec:sc_spectral_functions}
The main virtue  of the aFRG flows based on the SC1L ansatz for the effective average action as described in the previous section is its self-consistency beyond the soft-momentum limit which is a substantial improvement over previous LPA based aFRG calculations. It is nevertheless straightforward to implement, and the explicit analytic expressions provide a very clear and intuitive understanding of the underlying analytic structures.

These merits of the explicit SC1L ansatz at the same time clearly illustrate its limitations. Based on the explicit analytic expressions for self-energies of one-loop form, the approach is not able to resolve vertex corrections and $n$-particle thresholds with $n\ge 3$ at this lowest order for example. Including two-loop expressions for the self-energies together with one-loop structures in the 3-point vertex functions at the next order, on the other hand, the approach will quickly lose its simplicity.

Leaving the systematics of this expansion aside for now, in particular by not attempting to resolve the momentum-dependent substructures of three-point and higher $n$-point vertex functions, we can test at least the effects of including the higher orders in the self-energies. In order to achieve that, in this section we demonstrate how the numerical self-energies can alternatively be obtained in a fully self-consistent calculation beyond loop-expansion. 
In a first exploratory study we will use this fully self-consistent framework here to calculate the same $O(4)$ model spectral functions as above for comparison.

The main difference between the SC1L ansatz and the fully self-consistent calculation here will be the feedback of the full sigma two-point function reflecting its character as a broad two-pion resonance. As a result, the two-particle decay threshold $\pi^*\to\sigma +\pi $ described above will then become a broader three-particle resonance decay  $\pi^*\to\sigma +\pi \to 3\pi $. Apart from this important further conceptual improvement, however, the overall differences will turn out to be otherwise surprisingly small. Our comparison here might therefore lend further support to continue using the considerably simpler SC1L flows in more realistic effective field theory calculations for technical reasons in the future. 

\subsection{Self-consistent Euclidean flows}
\label{sec:sc_theory}
The fully self-consistent aFRG setup is based on reinserting the full $k$-dependent two-point functions into the flow of the effective potential. For the Euclidean system of coupled flow equations for effective potential and two-point functions this is rather straightforward. Based on the Euclidean results as input, the analytically continued flows will then be solved in a second step as described in the next subsection. 

First we briefly describe how to solve the Euclidean system consisting of flow equations for $\Gamma^{(2)}_{\pi,k}$, $ \Gamma^{(2)}_{\sigma,k}$ and for the effective potential $U_k$; the vertices $\Gamma^{(3)}_k$ and $\Gamma^{(4)}_k$ are extracted momentum independently from the effective potential as before. 
The resulting system of Euclidean flow equations then is of the following schematic form:
\begin{alignat}{1}
&\partial_k U_k[\phi]=\mathrm{Flow}_U\left(R_k,\Gamma^{(2),E}_{\pi,k},\Gamma^{(2),E}_{\sigma,k} \right),\label{eq:sc.euclidean_system}\\
&\partial_k \Gamma^{(2),E}_{\pi,k}[\phi=\phi_0; p\,]=\nonumber\\
&\qquad \qquad\left. \mathrm{Flow}_{\Gamma^{(2)}_\pi}\left(R_k, \Gamma_k^{(3)}, \Gamma_k^{(4)},\Gamma^{(2),E}_{\pi,k},\Gamma^{(2),E}_{\sigma,k} \right)\right|_{\phi=\phi_0},\nonumber\\
&\partial_k \Gamma^{(2),E}_{\sigma,k}[\phi=\phi_0;p\,]=\nonumber\\
&\qquad \qquad\left. \mathrm{Flow}_{\Gamma^{(2)}_\sigma} \left(R_k, \Gamma_k^{(3)}, \Gamma_k^{(4)},\Gamma^{(2),E}_{\pi,k},\Gamma^{(2),E}_{\sigma,k} \right)\right|_{\phi=\phi_0}.
\nonumber
\end{alignat}
As before, the flow of the effective potential is solved on a grid in field direction, whereas the flow equations of the two-point functions $\Gamma^{(2),E}_{\pi,k}$ and $\Gamma^{(2),E}_{\sigma,k}$ are evaluated at the physical minimum in the IR $\phi_0 =(\sigma_0,\vec \pi =0) $. As a further simplification we assume the field dependence of the two-point functions to be of the following form,
\begin{align}
\Gamma^{(2),E}[\phi,p]=\Gamma^{(2),E}[\phi_0,p]-\Gamma^{(2),E}[\phi_0,0]+m_E^2[\phi],
\label{eq:sc.lehmann}
\end{align}
i.e.~we split the two-point functions into field-independent but momentum-dependent parts plus a fully field-dependent but momentum-independent term which is identified with the Euclidean curvature mass extracted from the effective potential. This separation ansatz seems rather natural in view of the standard LPA or LPA$^\prime$ truncation, cf.~e.g.~\cite{Strodthoff:2016pxx}, but of course contains fully back-coupled objects calculated self-consistently here.

The flow equations for the two-point functions contain contributions analogous to those in Eqs.~(\ref{eq:Gamma2_sigma})-(\ref{eq:Gamma2_pion}). The expressions for the loop functions $I_k$ and the momentum-dependent $J_k(p) $ are of the same form as in App.~\ref{app:loop_functions}. 
Since the loop momentum integrations cannot be performed analytically here anymore, we set up a momentum-space grid for the Euclidean two-point functions and solve all momentum dependent expressions on every grid-point
at every scale $k$. In order to be ready for a straightforward extension to finite temperature in the future, in our numerical implementation we have introduced discrete explicit Matsubara sums for $p^0_n=2\pi n T $ at a small but finite Temperature of $T=10$~MeV, and a separate rotationally invariant grid for the magnitude of the spatial momentum variable $|\vec p|$ using an extrapolation for large momenta outside the grid, if necessary. As regulator function we use the standard three-dimensional Litim regulator, cf.~Eq.~(\ref{eq:regulator_3d}). We have checked explicitly that the number of Matsubara modes we included ($n_\text{max}\approx \pm 100$) are sufficient, and that the specific extrapolation procedure does not affect the results. 

For the initial conditions at the UV scale $\Lambda=500$~MeV we use Eq.~(\ref{eq:UV_pot}) for the effective potential with $b_1 =-0.2984\, \Lambda^2$ and $b_2=3.65$, and Eqs.~(\ref{eq:UV_Gamma2})-(\ref{eq:UV_Gamma2_2}) for the initial two-point functions of pion and sigma meson, with Eqs.~(\ref{eq:curvature_masses})-(\ref{eq:curvature_masses_2}) for the corresponding initial values of the field dependent $m_E^2[\phi]$. And as before, we add the explicit symmetry breaking term $c\sigma$ with  $c =0.014\, \Lambda^3$ to the resulting effective potential at the IR cutoff scale. The parameters $b_1$, $b_2$ and $c$ are again chosen to more or less reproduce the same phenomenological values for pion pole mass, with $m_\pi^p = 135.5$~MeV, and decay constant, with $ f_\pi\equiv  \sigma_0 = 93$~MeV.\footnote{Because of the small (few percent) effect, we did not bother to account for the renormalization according to Eq.~(\ref{eq:renormalized_minimum}) here.} Also the pole mass of the sigma meson then results in the same range as from the previous calculations. Since the numerical effort of further fine-tuning of the UV initial parameters gets unreasonably expensive in the fully self-consistent calculation, we are content here with the infrared values summarized in Tab.~\ref{tab:sf.observables}. For the purpose of comparison of the resulting self-consistent spectral functions with the considerably cheaper LPA and SC1L results these IR values are sufficiently close to those in Tab.~\ref{tab:pot_params}.

 \begin{table}[t!]
  \begin{center}
\begin{tabular}{ccccc}\hline
   $f_\pi $&$m_\pi^p$&$m_\sigma^p$&$m_\pi^c$&$m_\sigma^c$ \\ \hline \hline
   		93.0\text{ MeV}  &135.5\text{ MeV} & 353.7\text{ MeV}&137.4\text{ MeV}& 429.6\text{ MeV}\\  \hline
 \end{tabular}
 \end{center}
 	\caption{Values for $f_\pi \equiv \sigma_0$, the Euclidean curvature and the corresponding pole masses in the vacuum at $T=10$ MeV, as obtained with the fully self-consistent aFRG setup.}
	\label{tab:sf.observables} 
	 \end{table}
\subsection{Self-consistent aFRG flows}

Using the Euclidean results as input, we solve the analytically continued flow equations for real and imaginary parts of the (retarded) two-point functions of pion and sigma meson self-consistently in the next step. 
Because we have already calculated the field-dependent but momentum independent constants $m_E^2[\phi]$,    
the remaining system consists only of the momentum-dependent contributions to the flow. The corresponding  system of flow equations in the Minkowski spacetime is obtained from its Euclidean counterpart by applying the same analytic continuation procedure as above, cf.~Eq.~(\ref{eq:continuation}), in combination with Källén–Lehmann spectral representations for regulated propagators. The finite-temperature extension of the analytic continuation procedure is then also straightforward and parallels that used for the LPA based aFRG flows \cite{Tripolt:2013jra}.

In order to analytically continue the system of FRG flow equations for the retarded two-point functions, 
we first consider Källén–Lehmann spectral representations for the scale-dependent regulated Euclidean propagators with scale dependent spectral functions, which read 
\begin{align}
\label{eq.reconstruction}
D^{E}_{\alpha,k}(p_0,\vec{p})=
\int_{-\infty}^{\infty}\!\! d\omega\, \frac{\omega}{\omega^2+p_0^2}\,\, \rho_{\alpha,k}(\omega,\vec{p})\, ,
\end{align}
with $\alpha\in \{ \sigma,\pi\}$ for the different fields. 
The existence of such scale-dependent spectral representations follows from Cauchy's theorem and the use of three-dimensional regulator functions. 

Inserting these spectral representations of the propagators in the momentum-dependent loop functions $J_k(p)$ on the right-hand side of the flow equations, see Eq.~(\ref{eq:loop_functions_definitions_2}), these are of the following form,
\begin{align}
J_{\alpha\beta,k}(p_0,\vec{p})&=\SumInt_q  \int_{\omega_1,\omega_2,\omega_3}\hskip -.8cm d\omega_1 d\omega_2 d\omega_3 \hskip .2cm \partial_k R_k(\vec{q})  
\label{eq:sc.mink_J}\\
&\hspace{-1cm}\times\frac{\omega_1 \omega_2 \omega_3 \; \rho_{\alpha,k}(\omega_1,\vec{q})\rho_{\alpha,k}(\omega_2,\vec{q}) \rho_{\beta,k}(\omega_3,\vec{p}-\vec{q})}{\left(\omega_1^2+q_{0,n}^2\right) \left(\omega_2^2+q_{0,n}^2\right) \left(\omega_3^2+(p_0-q_{0,n})^2\right)}\, .
\nonumber
\end{align}
In these expressions we can then perform the Matsubara sum over $q_{0,n}=2\pi T n$ analytically and simply apply the standard analytic continuation procedure for the external $p_0$, cf.~Eq.~(\ref{eq:continuation}), after employing the $2 \pi T $ periodicity of the resulting statistical distribution functions \cite{Tripolt:2013jra}. 

The loop functions in (\ref{eq:sc.mink_J}) then obviously still involve the spatial momentum
integration. A major simplification thereby occurs, if we assume the scale-dependent spectral functions to be Lorentz invariant as they should be at zero temperature, i.e.
\begin{align} 
\rho_k(\omega, \vec{p}) &= \sign(\omega) \theta(p^2) \, \rho_k(p^2) \, . 
\end{align}
We can then substitute 
$ s=p^2=\omega^2-\vec p^{\,2}$ 
which amounts to using the zero-temperature spectral representation
\begin{align}
\label{eq:LehmannT0}
D^{E}_{\alpha,k}(p)=
\int_{0}^{\infty}\!\! ds\; \frac{\rho_{\alpha,k}(s)}{p^2+ s} 
\end{align}
in (\ref{eq:sc.mink_J}).
Note that this  is strictly speaking only valid at vanishing temperature, and for regulator functions that do not break the corresponding Euclidean $O(4)$ invariance. The latter is of course not the case for the three-dimensional regulators considered here, but the effect of the breaking of boost invariance induced by this was assessed already in\cite{Kamikado:2013sia} and found to be negligible as long as the magnitude of the spatial momentum stays well below the UV cutoff scale $\Lambda $. Also at finite temperature this will hold at best  approximately, if the temperature is sufficiently low. Collisional broadening is not included in the spectral functions in this way \cite{Bros:1992ey}, for example. We therefore assume Lorentz invariance to hold at least approximately here, in order to reduce the number of momentum arguments that have to be taken into account to resolve the momentum dependence of the two-point function.
Eq.~(\ref{eq:sc.mink_J}) then simplifies to
\begin{align}
J_{\alpha\beta,k}(p_0,\vec{p})&= \SumInt_q  \int_{s_1,s_2,s_3}\hskip -.4cm ds_1 ds_2 ds_3 \hskip .2cm \partial_k R_k(\vec{q})  
\nonumber \\
&\hspace{-1.2cm}\times\frac{\rho_{\alpha,k}(s_1)}{\left(s_1 +q_{0,n}^2 + \vec q^{\,2} \right)}\, \frac{\rho_{\alpha,k}(s_2) }{ \left(s_2+q_{0,n}^2+\vec q^{\,2}\right)} \label{eq:sc.mink_Js}\\
&\times \frac{\rho_{\beta,k}(s_3)}{\left(s_3+(p_0-q_{0,n})^2+(\vec p-\vec q)^{2}\right)} \, .
\nonumber
\end{align}
Here, we can in principle now also carry out the remaining spatial momentum integration and are hence left with the three spectral integrals over the invariant  $s_i\in[0,\infty)$ variables for every propagator involved in the loops on the right-hand side of the flow equations. These have to be performed numerically in every integration step.

The analytic continuation can then proceed in the standard way as described, and the originally Euclidean system of flow equations then turns into aFRG flows for the real and imaginary parts of the retarded two-point functions $\Gamma^{(2),R}_{\alpha ,k}(\omega) =-\Gamma^{(2)}_{\alpha ,k}(-i\omega^+)$ in Minkowski spacetime. We use $\vec p =0$ here and hence obtain complex aFRG flow equations of the following form 
\begin{align}
\partial_k {\Gamma^{(2),R}_{\pi,k}}(\omega) &= \label{eq:sc.minkowski_system}\\
& \hskip -.4cm
\mathrm{Flow}_{\Gamma^{(2)}_\pi} \left(R_k, \Gamma_k^{(3)}[\phi_0] \, , \Gamma_k^{(4)}[\phi_0];\rho_{\pi,k},\rho_{\sigma,k}\right),\nonumber\\
\partial_k {\Gamma^{(2),R}_{\sigma,k}}(\omega) &= \nonumber\\
& \hskip -.5cm
\mathrm{Flow}_{\Gamma^{(2)}_\sigma}\left(R_k, \Gamma_k^{(3)}[\phi_0], \Gamma_k^{(4)}[\phi_0];\rho_{\pi,k},\rho_{\sigma,k}\right) \, .\nonumber
\end{align}
The $k$-dependent spectral function, here with notation $\rho_k(\omega)\equiv \rho_k(\omega,0) $, are thereby recomputed at every integration step of the scale $k$ from
\begin{align}
\rho_{\alpha,k}(\omega)=\frac{1}{\pi}\frac{\Imag\Gamma_{\alpha,k}^{(2),R}(\omega)}{\left(\Real\Gamma_{\alpha,k}^{(2),R}(\omega)-k^2\right)^2+\left(\Imag\Gamma_{\alpha,k}^{(2),R}(\omega)\right)^2}\,,
\end{align}
for the pion and sigma meson respectively, where the form of the shift in the real part is due to the inclusion of the three-dimensional Litim regulator (\ref{eq:regulator_3d}) and the continuation procedure, which throws in a global minus sign from the definition of the two-point functions.

To further reduce the numerical effort, we further consider the product (in momentum space) of the identical propagators in the loop functions, 
\begin{align}
(D D)^{E,\text{eff}}_k(p_0,\vec{p})=D^{E}_k(p_0,\vec{p})\cdot D^{E}_k(p_0,\vec{p})\,,
\end{align}
and assume the same form of spectral representation to exist for this product also. Since both propagators are analytic functions in the cut-complex $p^2$ plane, the product is as well. Positivity of the discontinuity along the cut is not required here. The effective spectral function $\rho^\mathrm{eff}_k$ of the product then reads (again for the three-dimensional sharp regulator and vanishing spatial momentum)
\begin{align}
\rho^{\text{eff}}_k(\omega)&= \\
&\hskip -.5cm
-\frac{1}{\pi}\frac{2\;\Imag\Gamma_k^{(2),R}(\omega)~\left(\Real\Gamma_k^{(2),R}(\omega)-k^2\right)}{\left(\left(\Real\Gamma_k^{(2),R}(\omega)-k^2\right)^2+\left(\Imag\Gamma_k^{(2),R}(\omega)\right)^2\right)^2}\,.\nonumber
\end{align}
In this way we can further reduce the number of spectral integrals in
Eq.~(\ref{eq:sc.mink_J}) from three down to two.
The effects of using such a spectral representation for products of propagators should certainly be tested more carefully in the future. Our first results here are very promising, however, as we demonstrate below. 

The form of the loop functions $J_k(p_0)$ in (\ref{eq:sc.minkowski_system}), cf.\ Eq.~(\ref{eq:loop_functions_definitions_2}), at vanishing spatial momentum and $T=0$ for example then simplify as follows,
\begin{align}
J_{k}(p_0)&= \frac{k}{\pi^2}  \int_{s_1,s_2>0}\hskip -.8cm ds_1 ds_2 \hskip .1cm  \rho_{k}^\mathrm{eff}(s_1) \rho_k(s_2) \, \int_{-\infty}^\infty \frac{dq_0}{2\pi} 
\label{eq:spec_J2d}\\
&\hspace{-1cm}\times\frac{1}{s_1-s_2-p_0^2+ 2p_0 q_0} 
\Bigg\{ \sqrt{s_1+q_0^2} \arctan\Bigg(\frac{k}{\sqrt{s_1+q_0^2}}\Bigg)  \nonumber\\
& \hskip -.1cm - \sqrt{s_2+(q_0-p_0)^2} \arctan\Bigg(\frac{k}{\sqrt{s_2+(q_0-p_0)^2}}\Bigg) \Bigg\}
 \, .
\nonumber
\end{align}
The remaining $q_0$ integral can in principle also be calculated analytically, but the resulting expression is lengthy and not particularly illuminating. Most importantly, it still is an analytic function in the cut-complex $p_0^2$ plane and can thus be analytically continued in the external frequency via Eq.~(\ref{eq:continuation}) as needed for our aFRG approach. The explicit expression of the resulting $J_k^R(\omega)\equiv  J_k(-i\omega^+) $ is given in App.~\ref{app:C}. 

In the numerical calculation, the system of equations given by Eqs.~(\ref{eq:sc.minkowski_system}) is solved on a grid in frequency $\omega $.
In practice, it is therefore best to precompute the kernel of the spectral integrals in (\ref{eq:spec_J2d}) and store it as a fixed matrix on this frequency grid,
from which the two-dimensional spectral integrals of the form (\ref{eq:spec_J2d}) are then evaluated numerically in every $k$ step. 
The necessary cutoff $\Lambda_\omega$ in these spectral integrals determines the size of the $\omega$ grid and is chosen much larger than the RG cutoff scale $\Lambda$. We moreover extrapolate the spectral functions for $\omega>\Lambda_\omega$ with $\rho(\omega)\propto 1/\omega^2$ and verified explicitly that the results do not depend on the specific value used for this frequency cutoff  $\Lambda_\omega$. The initial conditions for the real and imaginary parts of the two-point functions at the UV scale $\Lambda$ are chosen as in (\ref{eq:UV_Gamma2})-(\ref{eq:UV_Gamma2_2}), here with a small but finite $\varepsilon$, and Euclidean curvature masses as extracted from the effective potential, 
\begin{alignat}{2}
&\Gamma^{(2),R}_{\pi,\Lambda}(\omega)&&=(\omega+{i}\varepsilon)^2-(m_{\pi,\Lambda}^c)^2\,,\\
&\Gamma^{(2),R}_{\sigma,\Lambda}(\omega)&&=(\omega+{i}\varepsilon)^2-(m_{\sigma,\Lambda}^c)^2 \, .
\end{alignat}

\begin{figure}[t]
	\includegraphics[width=0.45\textwidth]{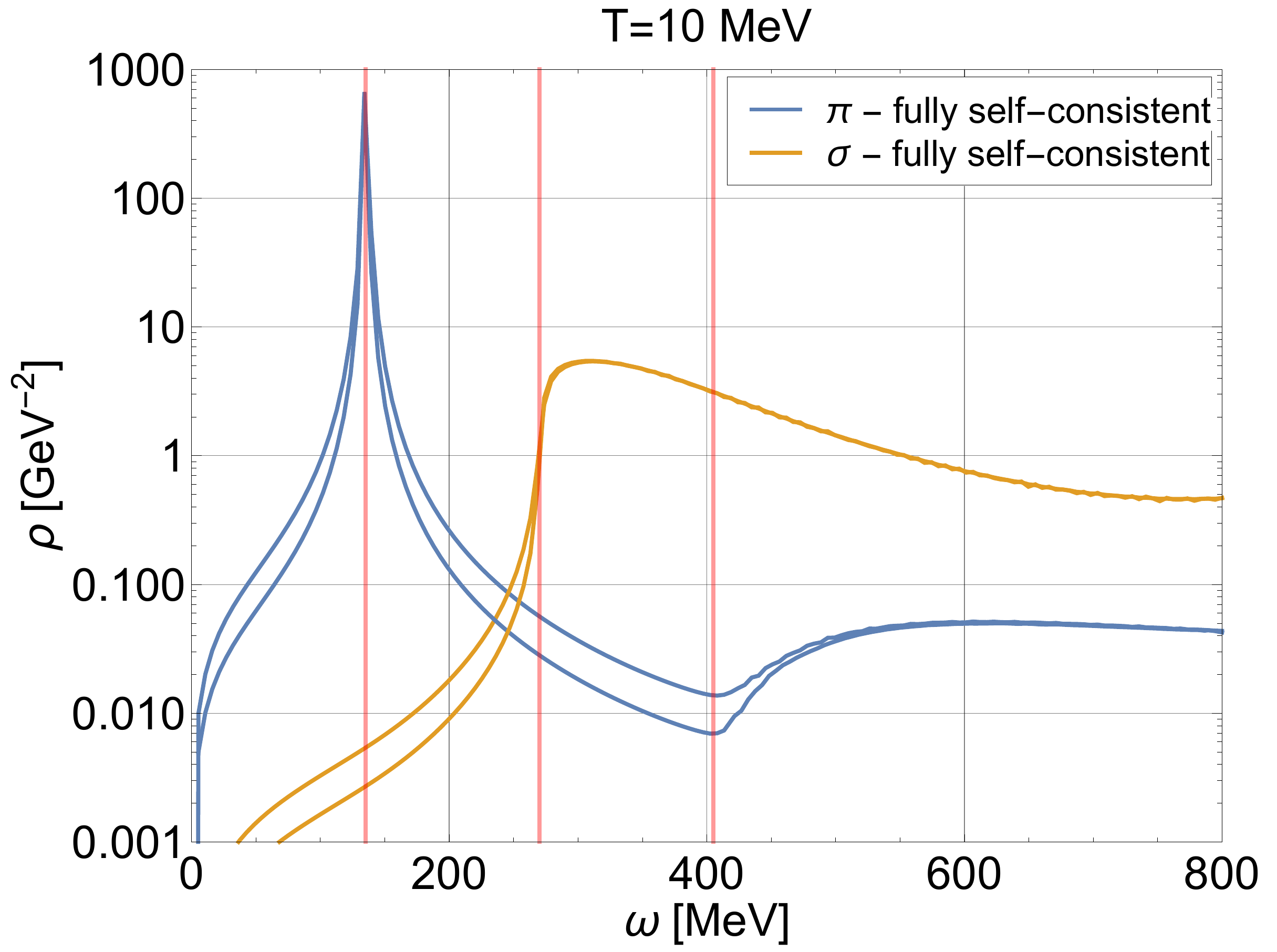}
	\caption{Pion and sigma meson spectral functions at vanishing spatial momenta over frequency $\omega$ obtained from the fully self-consistent aFRG flows at a temperature of $T=10$~MeV, from \cite{Jung:2019jlh}. The upper curves correspond to $\varepsilon=1$~MeV and the lower curves to $\varepsilon=0.5$~MeV. The red vertical lines indicate  the pion pole mass $m_\pi^p$ and multiples thereof for the $2\pi$ and $3\pi$ thresholds in sigma and pion, respectively.}
	\label{fig:sc_T10}
\end{figure}

\subsection{Self-consistent aFRG results and comparison}
\label{sec:sc_results}

Our first results for the $O(4)$ model spectral functions from the fully self-consistent aFRG flows, taken from \cite{Jung:2019jlh}, are shown in Fig.~\ref{fig:sc_T10}. These are based on the spectral representations as explained in the previous subsection. The discretization of the frequency for the Euclidean input here corresponds to a small residual temperature of \mbox{$T=10$ MeV} as mentioned above.

Because distributions are generally difficult to integrate numerically, we have to keep the imaginary parts of the frequencies in all retarded correlations positive with a small but finite $\varepsilon $ here. More sophisticated principal value prescriptions are needed to be able to take the limit $\varepsilon \to 0 $ analytically in spectral-representation based aFRG flows before they are solved numerically \cite{Wink:2020tnu}.

In this first exploratory study we simply stay off the real frequency axis a little bit. Smaller values of $\varepsilon $ require finer frequency grids for the analytically continued aFRG flows, and hence increased numerical costs. 
In order to exemplify the influence of these residual imaginary parts, the results of two such calculations with different values of $\varepsilon = 1$~MeV and  $\varepsilon = 0.5$~MeV are shown in Fig.~\ref{fig:sc_T10}. 

The vertical red lines in Fig.~\ref{fig:sc_T10} thereby indicate the location of the pion pole mass (at \mbox{$\omega=m_\pi^p\approx 135.5$ MeV}) in the pion spectral function and multiples thereof: (i) twice that for the two-pion threshold in the sigma spectral function (i.e.~at \mbox{$\omega=2m_\pi^p\approx 271$ MeV}); and, most significantly for the first time here, (ii) also three times that for the three-pion threshold (i.e.~at \mbox{$\omega=3m_\pi^p\approx 406.5$ MeV}) in the pion spectral function again. 

While the sigma meson spectral function shows the usual pronounced two-pion threshold at the right location of twice the pion pole mass, as it did already in our SC1L flows above, the figure also nicely illustrates the main effect of the full self-consistency here: towards smaller values of $\varepsilon$ the threshold in the pion spectral function nicely settles at $\omega=3m_\pi^p$ as it should, because the self-consistent sigma correlations in the flow now for the first time represent the broad two-pion resonance that the sigma should be. As a result, they contribute to the three-particle resonance decay \mbox{$\pi^*\rightarrow\sigma+\pi\to 3\pi$} which is impossible in any calculation based on one-loop structures as in our SC1L flows above. 

\begin{figure}[t]
	\includegraphics[width=0.45\textwidth]{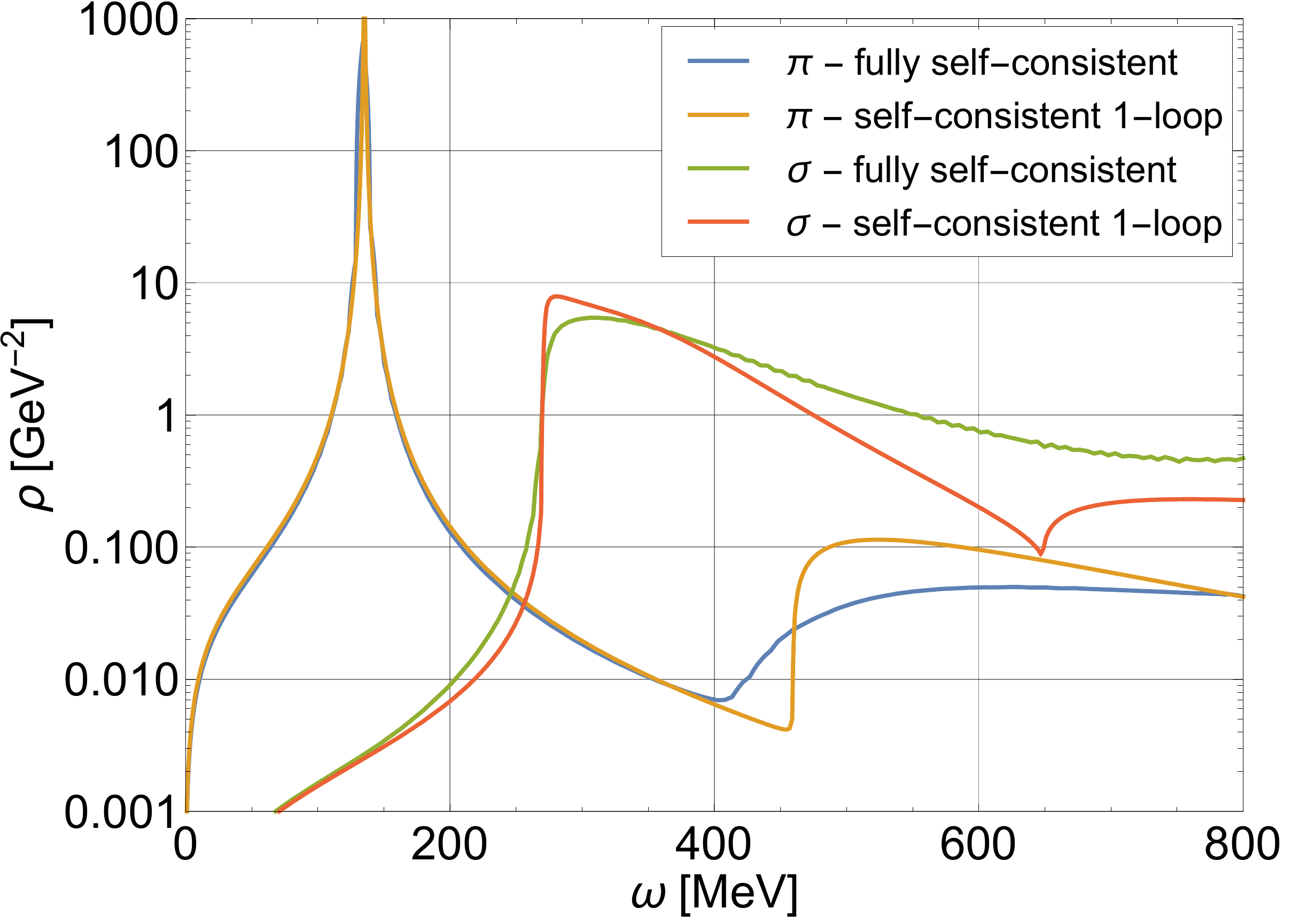}
	\caption{The pion and sigma spectral functions are shown vs.~external energy $\omega$ in the IR as obtained from the self-consistent one-loop setup and the fully self-consistent framework for $\varepsilon=0.5$~MeV.}
	\label{fig:spectral_sc}
\end{figure}

This important qualitative improvement comes at the price of an enormous increase in numerical complexity and costs, however. The SC1L aFRG flows of the previous sections are straightforward to implement and numerically very inexpensive compared to the fully self-consistent flows for the spectral functions, despite the simplifications we have used in this first study for the latter.\footnote{The fully self-consistent calculations were performed on a HPC cluster. Production runs with fine grids, large cutoff and small $\varepsilon $ took up to several thousand core hours while the SC1L calculations can typically be done within minutes on a single core.} We therefore close this section with a direct comparison of the fully self-consistent spectral functions with the corresponding ones from the self-consistent SC1L flows in Fig.~\ref{fig:spectral_sc}. In order to compare apples with apples instead of oranges we have recomputed the more economic SC1L results of Sec.~\ref{sec:2PF_and_SF} for this comparison once more, with the same value of $\varepsilon = 0.5$~MeV as used also in the fully self-consistent aFRG flows.

Although other parameters such as temperature and sigma pole mass for example do not match perfectly, compare~Tabs.~\ref{tab:IR_values} and \ref{tab:sf.observables}, the overall results of both schemes, with vastly different computational complexity, then look surprisingly similar. The particle peaks in the pion spectral functions are almost the same, with an only marginal relative shift due to the choice of parameters which lead to pion pole masses that differ by irrelevant $0.5$~MeV.

But even the decay threshold in the pion spectral function is not so dramatically different. 
While it starts at different values, i.e.~at  $\omega= m_\pi^p+m_\sigma^p \approx 460$~MeV in the SC1L result as compared to  \mbox{$\omega=3m_\pi^p\approx 406.5$ MeV} in the fully self-consistent calculation,
this qualitatively important shift is mainly due to a smearing of the unphysically sharp threshold observed in the self-consistent one-loop ansatz for the effective average action. 
The fully self-consistent aFRG calculation captures the broad nature of the 3-particle resonance decay $\pi\to \pi + \sigma \to 3 \pi $ certainly in a physically  more realistic manner than the 
SC1L calculation, where the resonance decay in the second step is missing, and which therefore produces an unrealistically sharp threshold for $\pi\to \pi + \sigma $. While it is reassuring to know that it is possible in principle to describe the three-particle resonance decay correctly, 
the question might arise whether the gain is worth the effort. Especially when one is interested in
heavy-ion collision observables such photons and dileptons for the electromagnetic spectral function, for example, which are measured from all stages of a collision and hence convoluted over the various temperatures and densities that occur during the evolution of the fireball, small differences such as those around threshold in the pion spectral function observed here will hardly be observable and will therefore not be relevant from a purely phenomenological perspective.


This kind of smearing of sharp structures is a general trend in the fully self-consistent and hence  more realistic calculation. As such, it is also observed in the sigma spectral function, where one can also clearly see that the threshold of the $\sigma^*\rightarrow \sigma+\sigma$ process (which corresponds to four pions in the final state now) gets smeared out completely and is no-longer visible in the fully self-consistent result. The sharp two-pion decay threshold on the other hand is located at the same energies again, owing to the self-consistency of both approaches and the nearly identical values for the pion pole masses in the two calculations. In conclusion, the comparison overall demonstrates how robust our aFRG results for spectral functions are at the present level.

\section{Summary and Outlook}
\label{sec:summary}
The main message of this work is the efficiency of the semi-analytic framework proposed here, which is referred to as the SC1L FRG flows above,
to calculate real-time correlation functions and spectral functions within the FRG from a self-consistent albeit still economic ansatz that includes parametrized self-energies directly in the effective average action. 

The explicit expressions for the self-energies of one-loop form in this approach have the correct domain of holomorphy of the two-point functions implemented by construction. The analytic continuation from Euclidean to Minkowski spacetime is therefore straightforward. The scale-dependent parameters in these self-energies are couplings and (real parts of resonance) pole masses. It is only these few parameters that have to be calculated self-consistently during the FRG flow which makes this scheme comparatively economic.  

We have compared our results for the spectral functions of pion and sigma meson in the $O(4)$ model obtained from these self-consistent SC1L flows first with our previous analytically continued aFRG flows based on the same local potential approximation (LPA) for the flows of effective potential and two-point functions. Because of this restriction, the self-consistency is then also restricted to the zero-momentum limit of the two-point functions obtained from the corresponding analytically continued aFRG flows. In particular, the non-trivial momentum dependence of these two-point functions is not fed back into the flows for effective potential and two-point functions, which results in a mismatch between the physical pole masses and the location of decay thresholds. This restriction is avoided in our new SC1L flows which are fully self-consistent. As an added bonus, the analytic structure of the two-point functions can be studied not only in the entire complex frequency plane but on all Riemann sheets. This is very illustrative and, in particular, it allows to identify the complex poles on unphysical Riemann sheets associated with resonances which are generally difficult to access numerically otherwise. 


In order to overcome the structural limitations due to the underlying one-loop form in the parametrizations of the SC1L self-energies, however, fully self-consistent aFRG calculations are necessary where the complete frequency and momentum dependence of the two-point functions is inserted back into the complete set of Euclidean and analytically continued flows. This is possible in principle by using spectral representations with scale-dependent spectral functions that are recomputed self-consistently in every numerical integration step of the FRG scale $k$. We demonstrate the  feasibility of this procedure here in an exploratory study of fully self-consistent $O(4)$ model spectral functions with several further simplifications for comparison. Our fully self-consistent results demonstrate that it is possible in principle to describe processes, from Cutkosky's cutting rule, that are beyond the one-loop structure of the previous approaches. Most significantly here, this is the three-particle resonance decay $\pi^*\to\sigma+\pi\to 3\pi $ which we observe at the correct three-pion threshold for the first time in this approach. The fully self-consistent aFRG scheme at this level is basically as good as it gets without including momentum dependent three and higher $n$-point vertex corrections.

Having said that, our comparison of the fully self-consistent aFRG with the SC1L results mentioned above also shows that this conceptually important improvement essentially arises by a smearing around the unphysically sharp threshold for $\pi^* \to\sigma+\pi $ of the latter. It might therefore quantitatively not be so significant in phenomenological applications, when there are other sources of smearing as well. Overall, our comparison can then serve to justify using the numerically considerably more economic and straightforward to implement SC1L framework which turned out to provide otherwise very robust results.  

The self-consistent SC1L approach can thus serve as a starting point for phenomenologically more realistic calculations of real-time quantities such as spectral functions and transport coefficients in a strongly interacting warm and dense medium. We therefore plan to extend this SC1L framework to finite temperature and density next, and to then apply it to effective hadronic theories for nuclear and neutron matter. This will eventually include also the electroweak response from calculating (axial-)vector and electroweak spectral functions of strong-interaction matter in the future.

\acknowledgments
Fruitful discussions and collaborations on the self-consistent aFRG calculations with Nils Strodthoff, Nicolas Wink and Jan Pawlowski are gratefully acknowledged. We also wish to thank Claudia H\"ohne for continued support and interest in this work. R.-A.~T.~is supported by the Austrian Science Fund (FWF) through Lise Meitner grant M 2908-N. This work was also supported by the Deutsche Forschungsgemeinschaft (DFG) through grant CRC-TR 211 ``Strong-interaction matter under extreme conditions'' and the German Federal Ministry of Education and Research (BMBF) through grant No.~05P18RGFCA. 

\appendix
\section{One-loop self energy}
\label{app:one_loop}
In this appendix we provide a derivation for the one-loop self energy as well as a discussion of its analytic properties. For a scalar particle, the Euclidean self-energy contribution from a single loop involving particles of different masses, $m$ and $M$, is given by the dimensionless, logarithmically divergent polarization integral, with $O(4)$ symmetric UV cutoff $\Lambda$,  
\begin{align}
\Pi_\Lambda(p^2) \equiv \frac{1}{2\pi^2}\int_\Lambda \! d^4q \, \frac{1}{M^2+\left(p - q\right)^2} \frac{1}{m^2+q^2},
\end{align}
where $q$ is the internal loop momentum, and $p$ the external momentum, see also Fig.~{\ref{fig:1L_diagrams}}. A prefactor of $1/2$ arises as the symmetry factor of the diagram, see e.g.~\cite{Das1997}, and is included here.

We can evaluate this integral analytically using spherical coordinates in four Euclidean dimensions,
\begin{align}
d^4q = 2\pi\: q^2 \sin^2\theta\: dq^2 \: d\theta\, .
\end{align}
With the substitution $\cos\theta = z$ we thus obtain
\begin{align}
      \Pi_\Lambda(p^2) =& \frac{1}{\pi} \int_{0}^{\Lambda^2}\!\! dq^2 \: \frac{q^2}{m^2+q^2} \, \times  \\ 
      & \hskip 1cm
        \int_{-1}^{1} \!\! dz\:\frac{\sqrt{1-z^2}}{M^2+p^2+q^2-2p \,q\,z} \, . \nonumber
\end{align}
Integrating $z$ then yields
\begin{align}
    \Pi_\Lambda (p^2) =& \frac{1}{4} \int_{0}^{\Lambda^2} \!\! dq^2 \; \frac{M^2+p^2+q^2}{p^2 \left(m^2+q^2\right)} \, \times  \\
    &\hskip 1cm \left(1-\sqrt{1-\frac{4p^2q^2}{\left(M^2+p^2+q^2\right)^2}}\right) \, .\nonumber
\end{align}
We are now interested in the renormalized self energy obtained from zero-momentum subtraction, $\Pi_R\left(p^2\right) = \Pi_\Lambda\left(p^2\right) - \Pi_\Lambda\left(0\right)$, which is then given by
\begin{widetext}
\begin{align}
\label{eq:PI_2masses_final}
\Pi_R\left(p^2\right) =& \frac{1}{2} \Bigg(
1 + \left( \frac{M^2+m^2}{M^2-m^2}
+ \frac{M^2-m^2}{p^2} \right) \ln\frac{M}{m}
-  \frac{\sqrt{\big(p^2 + (M+m)^2\big)\big(p^2 +(M- m)^2\big)}}{p^2} \, \times
 \\
&\hskip .7cm 
\bigg(\artanh\frac{p^2 + (M+m)(M-m)  }{\sqrt{\big(p^2 + (M+m)^2\big)\big(p^2 +(M- m)^2\big)}}+\artanh\frac{p^2 -(M+m)(M-m)}{\sqrt{\big(p^2 + (M+m)^2\big)\big(p^2 +(M- m)^2\big)}} \bigg)
\Bigg) \, . \nonumber 
\end{align}
\end{widetext}
If both particles in the loop have equal mass, $M=m$, the renormalized self energy simplifies to
\begin{align}
\label{eq:PI_1mass_final}
    \Pi_R(p^2) = 1
   - \sqrt{\frac{4m^2+p^2}{p^2}}\artanh \sqrt{\frac{p^2}{4m^2+p^2}} \; .
\end{align}
These renormalized expressions are used in Eqs.~(\ref{eq:1L_self_energies})-(\ref{eq:1L_self_energies_2}) for the calculation of sigma and pion self energies. 

The analytic continuation is performed by replacing the Euclidean external momentum by a real-time energy $\omega$ as discussed in Sec.~\ref{sec:LPA_setup}. 

Note that $\Pi_R(p^2)$ in Eq.~(\ref{eq:PI_2masses_final}) is an analytic function in the cut-complex $p^2$-plane as it must, from our zero-temperature one-loop calculation, here. It is therefore very well suited for parametrizing self-energies with the correct analyticity properties. The branch cut for timelike $-p^2 = s$ starts at $ s \ge (M+m)^2$ where the imaginary parts of both $\artanh$'s have the same sign and add up.

In contrast, for $0\le s \le (M-m)^2$ the two imaginary parts have opposite sign and cancel one another, while the apparent square-root cut for $(M-m)^2\le s \le (M+m)^2 $ is removed as long as the same Riemann sheet is used for the square-root in the prefactor as in the argument of $\artanh$ (with the identity $-i \artanh(iz) = \arctan(z) $).


Finally,  the $\artanh(z) $ has branch cuts along the real axis for $z \in (-\infty,-1]$ and $[1,\infty) $ with infinitely many Riemann sheets which are separated by multiples of $i\pi$. We can therefore add or subtract multiples of $i\pi$ to the $\artanh $ functions in Eqs.~(\ref{eq:PI_2masses_final})-(\ref{eq:PI_1mass_final}) in order to access different Riemann sheets. This allows to explore the self energies and hence the two-point functions on different Riemann sheets as well.
The second Riemann sheet for example is obtained by adding $+i\pi$ to all $\artanh $ functions in Eqs.~(\ref{eq:PI_2masses_final})-(\ref{eq:PI_1mass_final}), see also \cite{Tripolt:2016cya} for a similar discussion.

\section{LPA loop and vertex functions}
\label{app:loop_functions}
In this appendix we provide explicit expressions for the loop and vertex functions used in the flow equations for the two-point functions in the LPA setup, see also \cite{Kamikado:2013sia}. The loop functions are defined as
\begin{align}
\label{eq:loop_functions_definitions}
I^{(n)}_{\alpha,k}&=\int_q \partial_k R_k(q) D_{\alpha,k}^{n}(q)\,,\\
J_{\alpha\beta,k}(p)&=\int_q \partial_k R_k(q)
(D_{\alpha,k}(q))^2
D_{\beta,k}(q+p)\, , \label{eq:loop_functions_definitions_2}
\end{align}
with the scale-dependent Euclidean propagator
\begin{align}
D_{\alpha,k}(q)=\left(\Gamma_{\alpha\alpha,k}^{(2)}(q)+R_k(q)\right)^{-1},
\end{align}
and $\alpha \in \{ \sigma, \pi \}$. For the three-dimension Litim regulator, Eq.~(\ref{eq:regulator_3d}), we then get the following explicit expressions for $\vec{p}=0$,
\begin{align}
\label{eq:loop_functions}
I^{(2)}_{\alpha,k}&=\frac{k^4}{12\pi^2}\frac{1}{E_{\alpha,k}^3},\\
J_{\alpha\beta,k}(p)&=\frac{k^4}{12\pi^2}
\frac{(E_{\alpha,k}+E_{\beta,k})^2(2E_{\alpha,k}+E_{\beta,k})+E_{\beta,k}p_0^2}
{E_{\alpha,k}^3E_{\beta,k}(p_0^2+(E_{\alpha,k}+E_{\beta,k})^2)^2},
\end{align}

The four-point vertex functions can be obtained from the ansatz for the effective potential, Eq.~(\ref{eq:Ansatz_Gamma}), and are given explicitly by
\begin{align}
\label{eq:vertex_functions}
\Gamma^{(4)}_{\sigma\sigma\sigma\sigma,k}&=12U^{(2)}_k+48\phi^2U^{(3)}_k+16\phi^4U^{(4)}_k,\\
\Gamma^{(4)}_{\sigma\sigma\pi\pi,k}&=4U^{(2)}_k+8\phi^2U^{(3)}_k,\\
\Gamma^{(4)}_{\pi\pi\pi\pi,k}&=\sqrt{176/3}\:U^{(2)}_k,
\end{align}
where the last vertex function contains contributions from the case where the internal pion is the same as the external one as well as from the other case where they are different.

\section{Spectral integral for loop functions}
\label{app:C}

The loop functions of the form in Eq.~(\ref{eq:spec_J2d}) for the three-dimensional Litim regulator (\ref{eq:regulator_3d}) can be written as follows, 
\begin{align}
J_{k}(p_0)&= \frac{k}{\pi^2}  \int_{s_1,s_2>0}\hskip -.8cm ds_1 ds_2 \hskip .1cm  \rho_{k}^\mathrm{eff}(s_1) \rho_k(s_2) \; K_k^J(s_1,s_2,-p_0^2) \, .
\end{align}
The dimensionless kernel $K_k^J(s_1,s_2,-p_0^2) $ in this double spectral integral  for $T=0$ can be read off from Eq.~(\ref{eq:spec_J2d}), which was obtained from first integrating the regulated spatial loop momentum (with $q=|\vec q|$ here) in
\begin{align}
 K_k^J(s_1,s_2,-p_0^2) &= \int_{-\infty}^\infty \frac{dq_0}{2\pi} \int_0^k q^2 dq \label{eq:spec_KJ}\\
 & \times \frac{1}{s_1+q_0^2+q^2} \, \frac{1}{s_2+(q_0-p_0)^2+q^2} \, .\nonumber
\end{align}
The analytic expression after integrating all loop variables, and after analytic continuation, is then given by
\begin{widetext}
\begin{align}
    K_k^J(s_1,s_2,\omega^2) &=
    \frac{1}{8} \, \Bigg\{ 
    \, \frac{(s_1-s_2)}{2\omega^2} \, \ln\bigg(\frac{s_2 ( k+\sqrt{k^2+s_1})^2 }{s_1 (k+\sqrt{k^2+s_2})^2 }\bigg) 
    + \frac{1}{2} \, \ln\bigg(\frac{( k+\sqrt{k^2+s_1})^2 (k+\sqrt{k^2+s_2})^2}{s_1  s_2}\bigg) \label{eq:sec_K_J}\\
    &\hskip -.1cm - \frac{\sqrt{\omega^4-2(s_1+s_2)\omega^2 + (s_1-s_2)^2}}{\omega^2} \, \Bigg( \,\artanh\bigg(\frac{k(s_1-s_2+\omega^2)}{\sqrt{(k^2+s_1)
    (\omega^4-2(s_1+s_2)\omega^2 + (s_1-s_2)^2})}\bigg) \nonumber\\
    & \hskip 6cm + \artanh\bigg(\frac{k(s_2-s_1+\omega^2)}{\sqrt{(k^2+s_2)
    (\omega^4-2(s_1+s_2)\omega^2 + (s_1-s_2)^2})}\bigg)\Bigg)\Bigg\} \, . \nonumber
\end{align}
\end{widetext}
The structural similarity with the analytic expression of the self-energy in Eq.~(\ref{eq:PI_2masses_final}) is of course no coincidence. The discussion of the analytic structure follows the same line of arguments. In particular, it also has branch cuts along the real axis for $\omega^2\ge (\sqrt{s_1} + \sqrt{s_2})^2 $.

The analytically continued form of the retarded loop function $J_k^R(\omega)\equiv  J_k(-i\omega^+) $  in Eq.~(\ref{eq:spec_J2d}) can then be written as follows,
\begin{align}
J_{k}^R(\omega)&= \frac{k}{\pi^2}  \int_{s_1,s_2>0}\hskip -.8cm ds_1 ds_2 \hskip .1cm  \rho_{k}^\mathrm{eff}(s_1) \rho_k(s_2) \; K_k^J(s_1,s_2,{\omega^+}^2) \, .
\end{align}

Finally, for $k^2<s_1,s_2$, a very good further approximation of $K_k^J(s_1,s_2,\omega^2)$ is obtained from expanding
\begin{align}
    K_k^J(s_1,s_2,\omega^2) &= \frac{k}{6} \,\frac{\sqrt{s_1}+\sqrt{s_2}}{\sqrt{s_1}\sqrt{s_2}}\\
    & \hskip -1cm \times \left(\frac{k^2}{s_1+s_2 +2 \sqrt{s_1}\sqrt{s_2} -\omega^2}
    \, +\,\mathcal O\Big(\frac{k^4}{s_1 s_2}\Big) \right) \, .
    \nonumber
\end{align}
The leading term of this expansion was actually used for simplicity to produce the results presented in Sec.~\ref{sec:sc_results} above. The higher-order corrections affect the flow only close to the UV cutoff. These might in turn induce some corrections to the spectral functions of pion and sigma meson for high frequencies $\omega $, of the order of the UV cutoff $\Lambda $ as well, which are however not the focus of our comparison with the considerably more economic SC1L calculations here.

%

\end{document}